\pdfoutput=1
\documentclass[twocolumn]{article}

\usepackage{graphicx}
\usepackage{caption}
\usepackage{subfig}
\usepackage{authblk}

\usepackage[normalem]{ulem}

\newcommand{\graphwidth}{0.48\textwidth}
\newcommand{\halfgraphwidth}{0.24\textwidth}

\usepackage{algorithm2e}

\title{Hardware Support for Address Mapping in PGAS Languages; a UPC Case Study}

\author[1]{Olivier Serres \thanks{serres@gwu.edu}}
\author[2]{Abdullah Kayi \thanks{abdullah.kayi@intel.com}}
\author[1]{Ahmad Anbar \thanks{anbar@gwu.edu}}
\author[1]{Tarek El-Ghazawi \thanks{tarek@gwu.edu}}

\affil[1]{
ECE Department\\
The George Washington University\\
Washington DC, USA\\
}

\affil[2]{
Intel\\
Hillsboro, OR, USA\\
}

\begin{document}

\date{}
\maketitle

\thispagestyle{empty}

\begin{abstract}
The Partitioned Global Address Space (PGAS) programming model strikes a balance
between the locality-aware, but explicit, message-passing model (e.g. MPI) and
the easy-to-use, but locality-agnostic, shared memory model (e.g. OpenMP).
However, the PGAS rich memory model comes at a performance cost which can
hinder its potential for scalability and performance. To contain this overhead
and achieve full performance, compiler optimizations may not be sufficient and
manual optimizations are typically added. This, however, can severely limit
the productivity advantage. Such optimizations are usually targeted at reducing
address translation overheads for shared data structures. This paper proposes
a hardware architectural support for PGAS, which allows the processor to
efficiently handle shared addresses. This eliminates the need for such
hand-tuning, while maintaining the performance and productivity of PGAS
languages. We propose to avail this hardware support to compilers by
introducing new instructions to efficiently access and traverse the PGAS memory
space. A prototype compiler is realized by extending the Berkeley Unified
Parallel C (UPC) compiler. It allows unmodified code to use the new
instructions without the user intervention, thereby creating a real productive
programming environment. Two different implementations of the system are
realized: the first is implemented using the full system simulator Gem5, which
allows the evaluation of the performance gain. The second is implemented using
a softcore processor Leon3 on an FPGA to verify the implementability and to
parameterize the cost of the new hardware and its instructions. The new
instructions show promising results for the NAS Parallel Benchmarks implemented
in UPC. A speedup of up to 5.5x is demonstrated for unmodified and unoptimized
codes. Unoptimized code performance using this hardware was shown to also
surpass the performance of manually optimized code by up to 10\%.
\end{abstract}

\section{Introduction}

\sloppy
Considering parallel programming models, there has always been a trade-off
between programmability and performance.
For instance, the two widely accepted parallel programming models, shared
memory and message passing have both their own advantages and disadvantages.
The shared memory model gives the programmer an easy-to-program shared view of
the memory, that allows one-sided communication.
At the same time, the shared memory model has no notion of data locality, which
might lead to severe degradation in performance.
This degradation in performance is attributed to tasks generally working on
remote data, as there is no way to infer from a programmers perspective, which
data are local and which are remote to each task.
On the other hand, the message passing model allows the programmers to fully
exploit the data locality to achieve high performance at the cost of
productivity.
As this model allows only explicit two sided communication, it is the
programmer's responsibility to explicitly express the data movements, both at
the sender and receiver tasks.

Partitioned Global Address Space (PGAS) programming model is capturing the good
of these earlier models.
The PGAS programming model has shown a great potential for scalability and
performance, as it furnishes a partitioned memory view allowing the
programmer to exploit the data locality.
At the same time it still maintains the shared view of the memory, and thus
still maintains the productivity advantage.
However, this partitioned global view of the memory entails a more complex
addressing mode to map the programmer's view of the memory to the actual
physical layout of the memory.
This mismatch necessitates two major elements (1) a new way of representation of
shared addresses in the global view, and (2) the translation between the global
representation and the virtual address representation.
For example, in the Unified Parallel C (UPC) language, three different fields
are necessary to represent a shared address.
This creates a significant overhead within the runtime to translate this
representation to a regular memory address.
This definitely increases the programming complexity which in turn forces users
to manually optimize their codes, clearly reducing the productivity
advantage. Manual optimizations to improve performance usually includes complex
pointers and MPI like messaging which totally degrades the PGAS programming
productivity.

To solve this issue, we propose a hardware support mechanism to handle complex
PGAS address mapping tasks with hardware assistance via newly introduced
instructions.
This eliminates the need of manual tuning of the code to work around this
specific problem.
A PGAS compiler can make use of such instructions to efficiently translate from
the shared address representation to the actual memory address representation
on the target machine.
The proposed hardware support aims to close the performance gap between shared
pointer addressing and private pointer addressing with modest hardware changes.
To verify performance improvements and feasibility of our proposed hardware
mechanism, we conducted full-system simulation as well as FPGA prototyping.

The rest of this paper is organized as follows: Section~\ref{sec:upc} presents
the UPC language and particularly its memory model.
Section~\ref{sec:related} reviews the related work, Section~\ref{sec:pgas_hw}
discusses the proposed PGAS hardware support.
Section~\ref{sec:experimental_setup} describes our prototype implementations;
one is using the Gem5 full system simulator to obtain performance results
with up to 64 cores, the other one implement the hardware support in an FPGA
along with a softcore processor allowing us to evaluate the feasibility and
the chip area needed for such a design.
In Section~\ref{sec:results} we present and discuss the results.
Finally, Section~\ref{sec:conclusion} concludes the paper and considers future
work.

\section{Unified Parallel C}
\label{sec:upc}

\begin{figure}[!tbp]
    \centering
    \includegraphics[width=\graphwidth]{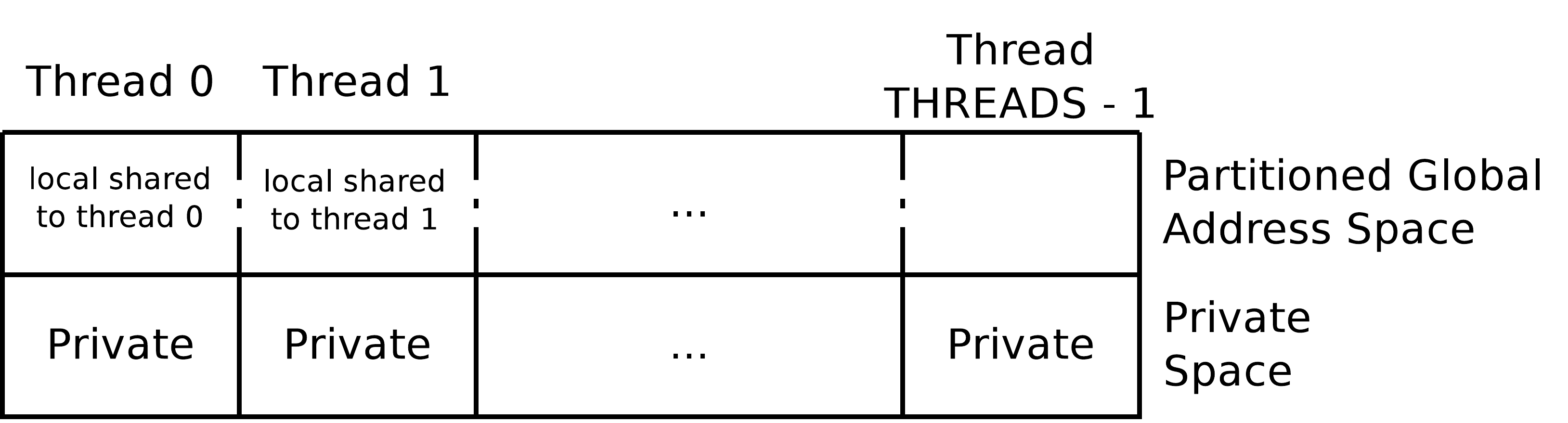}
    \caption{UPC memory model}
    \label{fig:upc_memory_model}
\end{figure}

\begin{figure}[!tbp]
    \centering
    \includegraphics[width=\graphwidth]{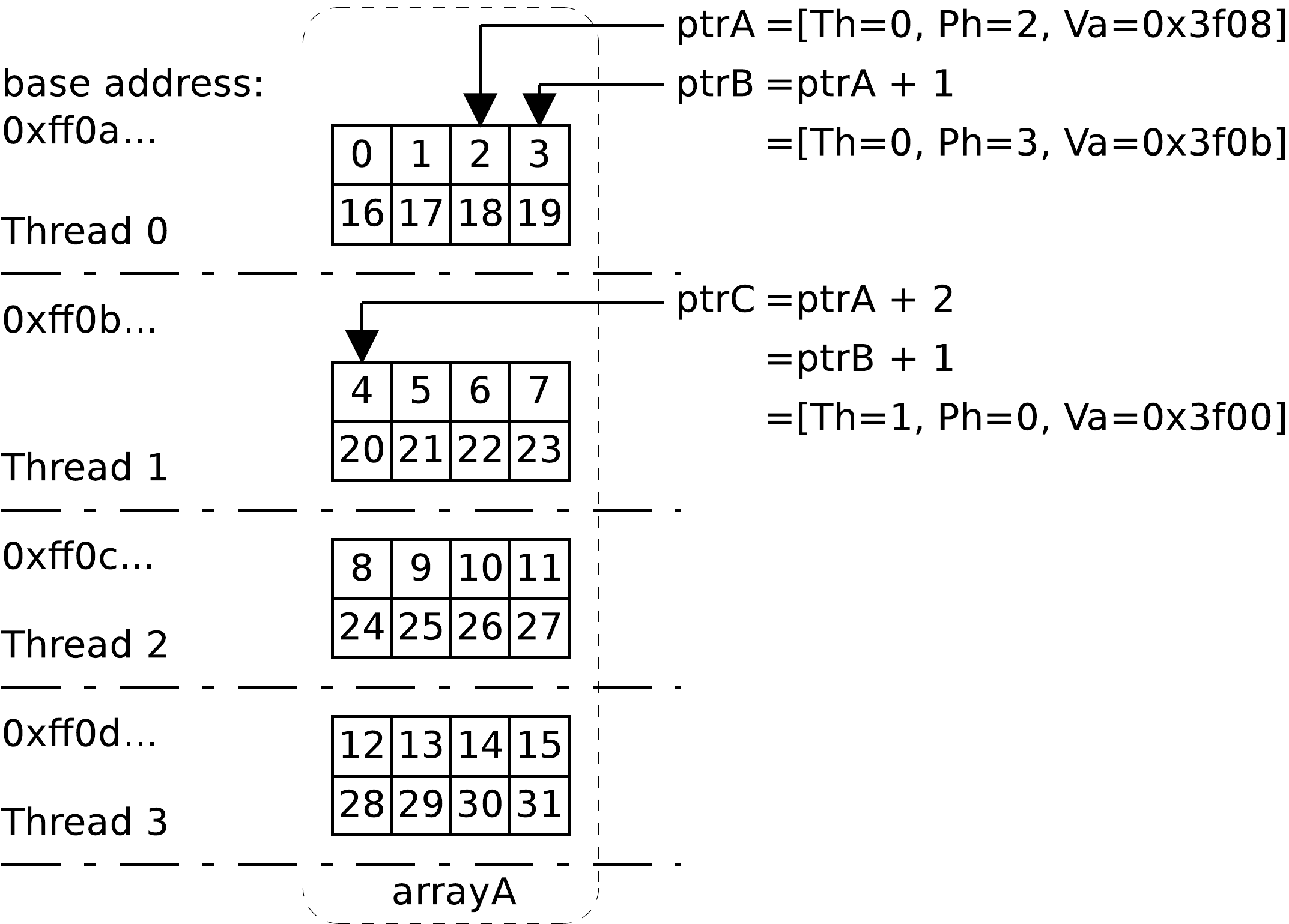}
    \caption{UPC Memory layout for \texttt{arrayA}, some shared pointers are also represented}
    \label{fig:memory_layout}
\end{figure}

Unified Parallel C (UPC) is a parallel extension of ISO C 99 programming
language implementing the PGAS model \cite{2005_upc_specifications_1.2}.
It follows an SPMD execution model in which a specified number of threads are
executed in parallel.
UPC realizes the PGAS memory model by providing a shared memory view across the
system that can be accessed by any thread; each thread having an affinity to
the part of the shared memory residing locally.
In addition, each thread has access to a private space that is accessible only
by the thread itself.
The private space has low overheads and allows for the
best performance.
Figure~\ref{fig:upc_memory_model} provides an overview
of the UPC memory model.
The language also provides all the facilities needed for parallel programming:
locks, memory barriers, collective operations, etc.
Accesses to either the shared or the private memory space are syntactically
identical and done through simple variable accesses or assignments.

The distribution of shared data across the different threads is controlled
by a block size specified by the user for a given array: elements are
distributed in block size elements in a round robin fashion.
Thus, the blocking factor gives the programmers a mechanism to control the data
distribution in the shared space.
For example, Figure~\ref{fig:memory_layout} presents how the elements of the
following array are distributed across 4 UPC threads.
Each thread has its own contiguous address space starting at \textit{base
address}.

\begin{verbatim}
shared [4] int arrayA[32];
\end{verbatim}

In order to address such arrays, a UPC shared pointer can be used.
Shared pointers are similar to C pointers but are able to traverse shared arrays
in their normal ordering.
They effectively provide a mapping from the logical array order to the actual
physical location of the data in the system across the whole shared space.

The shared pointer fields allow to perform the mapping between the shared space
and the physical distribution.
A shared pointer is usually composed of three elements: \textbf{thread:} thread
affinity of the pointed data, \textbf{virtual address:} address of the current
element in the local space and \textbf{phase:} position inside the current
block. This allows the programmer to traverse the array in a logical way.
Current implementations of UPC usually use 64 bits to represent a shared
pointer.
Even if a compiler uses its own internal representation, the UPC
specification \cite{2005_upc_specifications_1.2} provides the following
function to indirectly access them: upc\_threadof, upc\_phaseof,
upc\_addrfieldof and upc\_resetphase.  Figure~\ref{fig:memory_layout} also
presents a few examples of shared pointers (ptrA, ptrB and ptrC).

\section{Related Work}
\label{sec:related}

Previous studies have analyzed both the productivity advantage and the
performance of PGAS languages under different levels of manual code
optimizations and on a variety of systems.
Many of those studies have shown that the shared pointer arithmetic and the
address translation are the main performance impediments in UPC codes.

In \cite{2004_fc_ipdps_productivityanalysisofupc}, Cantonnet et al.
demonstrated clear advantage in terms of productivity when using UPC compared
to MPI.  Their experiments has shown that UPC has consistent improvement over
MPI in terms of number of lines of code, number of characters, and conceptual
effort to write the same program. Ebcioglu et al.
\cite{2006_ke_pphec_prodthreeproglangs} performed a 4.5 day study on 27
subjects, to compare the productivity of parallel programming languages.
In this study, they compared the time to reach correct output when using
several parallel languages including: C+MPI, UPC.
The study showed that the use of a PGAS language can improve the productivity.

Along with the productivity studies, many efforts were performed in the
direction of evaluating the potential of achieving performance using UPC.
In \cite{2007_ky_pasco_prodperfpgas}, the authors demonstrated that hand-tuned
UPC code can achieve comparable performance to, and sometimes even better than,
code in MPI.  \cite{2005_zz_ipdps_upcbench} evaluates the performance of
different UPC compilers on 3 different machines: a Linux x86 cluster, an
AlphaServer SC and a Cray T3E.

In \cite{2006_teg_futuregen_benchmarkingparallelcompilers},
El-Ghazawi et al. clearly demonstrated the overhead of the PGAS shared memory
model. They proposed a framework to assess the compilers and the runtime systems
capabilities to optimize such overheads.
In order to solve those issues, different compiler optimizations have been
researched including optimization techniques such as lookup tables:
\cite{2005_fc_ipdps_fasttranslation,
2011_os_ipdpsw_address_translation_opt_arrays}; the reduced overhead
is still significant and the methods can use a good amount of memory.
Also, alternative representations for shared pointers have been implemented,
for example phaseless pointers are used for shared addresses with a block size
of 1 or infinity \cite{2003_wyc_ics_perfberkeley}, this is only applicable to
a few cases and still present a significant overheads.

Multiple systems have implemented Hardware support for shared memory across a
system. For example, the T3D supercomputer used a 'Support Circuitry' chip
located between the processor and the local memory
\cite{1995_rha_acmarchnews_emperical_t3d_compiler}; this chip, on top of
providing functionality like message passing and synchronization, allowed the
processor to access any memory location across the machine.
In \cite{2010_hf_ipdpsw_hardwaresupportpgas} is proposed a network engine
especially designed for PGAS languages: it allows network communication
between nodes by mapping other nodes memory space accross the network and
providing a relaxed
memory consistency model best suited for PGAS.  Results were only presented in
terms of read/write throughput and transaction rates as no PGAS applications or
benchmarks were tested.  This approach is complementary to our work, as it
focused on the network interface for PGAS languages and this paper focuses on
the shared space addressing. Combining both an efficient addressing a an
efficient network interface would provide a very efficient support for
PGAS; this is noted as future work.

More interestingly, the T3E \cite{1996_sls_sigplannotices_sync_comm_t3e,
2000_mmm_ibuk_efficient_address_translation_t3e,
1999_wwc_ida_intro_upc_lang_spec} improved on that by providing
E-registers and a 'centrifuge' hardware allowing to perform some mapping for
arrays using 4 registers (index, mask, base address, stride and addend). This
provides a good support for the data layout of PGAS languages at the
level of the network interface, however
this approach has multiple drawbacks that needs to be addressed: it uses a
great number of registers, the registers are memory-mapped and hence relatively
slow to access and the hardware is outside of the processor chip, close
to the network interface making it useless to improve the performance of the
very frequent local accesses.

\section{PGAS hardware support}
\label{sec:pgas_hw}

\begin{algorithm}[tbp]
    \SetAlgoLined
    \SetKwInOut{Input}{input}\SetKwInOut{Output}{output}
    \Input{blocksize, elemsize, increment, numthreads, shptr}
    \Output{nshptr}

    phinc = shptr.phase + increment

    thinc = phinc / blocksize

    nshptr.phase = phinc \% blocksize

    blockinc = (shptr.thread + thinc) / numthreads;

    nshptr.thread = (shptr.thread + thinc) \% numthreads;

    eaddrinc = (nshptr.phase - shptr.phase) + blockinc * blocksize;

    nshptr.va = shptr.va + eaddrinc * elemsize;

    \caption{Shared pointer incrementation}
    \label{algo_ptr}
\end{algorithm}

In this section, we discuss the overheads of the current approach, the
general principles behind the hardware support and how the hardware
is made available to compilers by extending the instruction set.

\subsection{PGAS Memory Model Overheads}

Shared pointer manipulations are currently performed in software.
Incrementing a shared pointer consists of updating the three fields of the
pointer to point to a different element in the shared array.
This is done using algorithm similar to the one presented
in Algorithm~\ref{algo_ptr}.
The algorithm increments \texttt{shptr} to the new pointer \texttt{nshptr}.
This is a particularly complex operation involving additions, subtractions,
multiplications and divisions. It requires temporary registers which
can increase register spill.
It is to be noted that to increment a pointer, extra information like the array
block size, the element size (e.g. 4 bytes for int) and the number
of running threads are needed.

The shared pointer manipulations may be optimized by compilers in
some cases using various methods like using a
simpler representation for some kind of pointers or optimizing the shared
pointer manipulation away when accessing the local space. However, this is not
always feasible due to the complexity of the compiled code or due to
dependencies on other results or functions from a different compilation unit.

When an element is accessed, the shared pointer needs to be converted to a
virtual address that the processor can manipulate. This is not as compute
intensive but it still requires to lookup the base address for the thread
pointed to and perform an addition. This can greatly increase the time required
to access an element.

\subsection{PGAS Memory Model Hardware Support}

The hardware support proposed here adds a specific support for the shared
pointers.
This allows to close the performance gap between the shared space addressing
and the private space addressing.
At least two different types of operations need to be optimized in order to get
an efficient addressing of the shared space: incrementing the pointers allowing
to traverse arrays and translating the shared addresses to the final physical
address allowing for reading and writing. Other operations like testing for
the locality of a shared pointer (checking if a shared pointer points to local
data or not, which can be used to quickly call a communication sub-routine
if the data is off-node) can also benefit from the same hardware support.

Algorithm~\ref{algo_ptr} can be pipelined fairly well for a hardware
implementation.
This is especially true when considering the most common case of having
\texttt{numthreads}, \texttt{blocksize} and \texttt{elemsize} as powers of 2.
This assumption is used in our implementations, allowing to replace the
divisions and multiplications with simple shifting and masking.
At the same time, it does not restrict the user as the compiler can fall-back
on the software implementation for the cases not supported by the hardware.

When using a shared pointer to access data, the physical location should be
computed.
This is done by adding the \textit{base address} of the thread specified by the
pointer to the virtual address contained in the pointer.
The system virtual address will then be transformed to the final physical
address using the conventional translation lookaside buffer (TLB) hardware.

At least two different implementations are possible for the address translation:
the thread address spaces can be starting at regular intervals allowing the
\textit{base address} to be simply computed from the thread number (similarly
to \cite{2010_hf_ipdpsw_hardwaresupportpgas}), or a lookup table can be used to
retrieve the \textit{base address}.
The first method is more restrictive in the addresses possibly used but more
scalable as it does not require storing a table of base addresses.
We used the second one in our prototype implementation for simplicity.

For example, for \texttt{ptrC} of Figure~\ref{fig:memory_layout}, the system
virtual address of the element would
be computed by retrieving the base address of thread 1 and adding the virtual
address from the shared pointer: $0xff0b00000000 + 0x3f00 = 0xff0b00003f00$.

\subsection{Instruction Set Extension}

The new hardware can be availed to the compilers with new instructions:
instructions to manipulate pointers and instructions to load/store using
a shared pointer as an address.
Shared addresses can be stored in the normal processor registers, the
other needed information like the block size or the element size can
be directly encoded in the instructions.
The increment value can be an immediate value or coming from a register.
In the following implementations, we also used a special register to store the
number of UPC threads for the currently running program.

\section{Experimental Setup}
\label{sec:experimental_setup}

Our experimental setup is composed of two parts, one using full system
simulation allowing to evaluate the performance characteristics of
the hardware support, the other one is an hardware implementation
using FPGAs allowing us to study the implementation details and
evaluate the chip area needed for such an extension.

\subsection{Full System Simulation}

\begin{table}[!tbp]
    \centering
    \caption{Instructions Added to the Alpha ISA}
    \begin{tabular}{|l|l|}
        \hline
        \multicolumn{2}{|c|}{Shared Address Loads} \\
        \hline
        pgas\_ldbu & Load Byte Unsigned (8 bits)\\
        pgas\_ldwu & Load Word Unsigned (16 bits)\\
        pgas\_ldl  & Load Long Unsigned (32 bits)\\
        pgas\_ldq  & Load Quad Unsigned (64 bits)\\
        pgas\_lds  & Load S\_float (32 bits, float)\\
        pgas\_ldt  & Load T\_float (64 bits, double)\\
        \hline
        \multicolumn{2}{|c|}{Shared Address Stores} \\
        \hline
        pgas\_stb & Store Byte Unsigned (8 bits)\\
        pgas\_stw & Store Word Unsigned (16 bits)\\
        pgas\_stl & Store Long Unsigned (32 bits)\\
        pgas\_stq & Store Quad Unsigned (64 bits)\\
        pgas\_sts & Store S\_float (32 bits, float)\\
        pgas\_stt & Store T\_float (64 bits, double)\\
        \hline
        \multicolumn{2}{|c|}{Shared Address Incrementations} \\
        \hline
        pgas\_inc\_imm & Address increment, immediate\\
        pgas\_inc\_reg & Address increment, register\\
        \hline
        \multicolumn{2}{|c|}{Initialization} \\
        \hline
        set\_threads & Initialize the 'threads' register\\
        set\_base\_address & Set the base address\\
                           & look-up table\\
        \hline
    \end{tabular}
    \label{tab:isa_extension}
\end{table}

\begin{figure}[!tbp]
    \centering
    \includegraphics[width=0.43\textwidth]{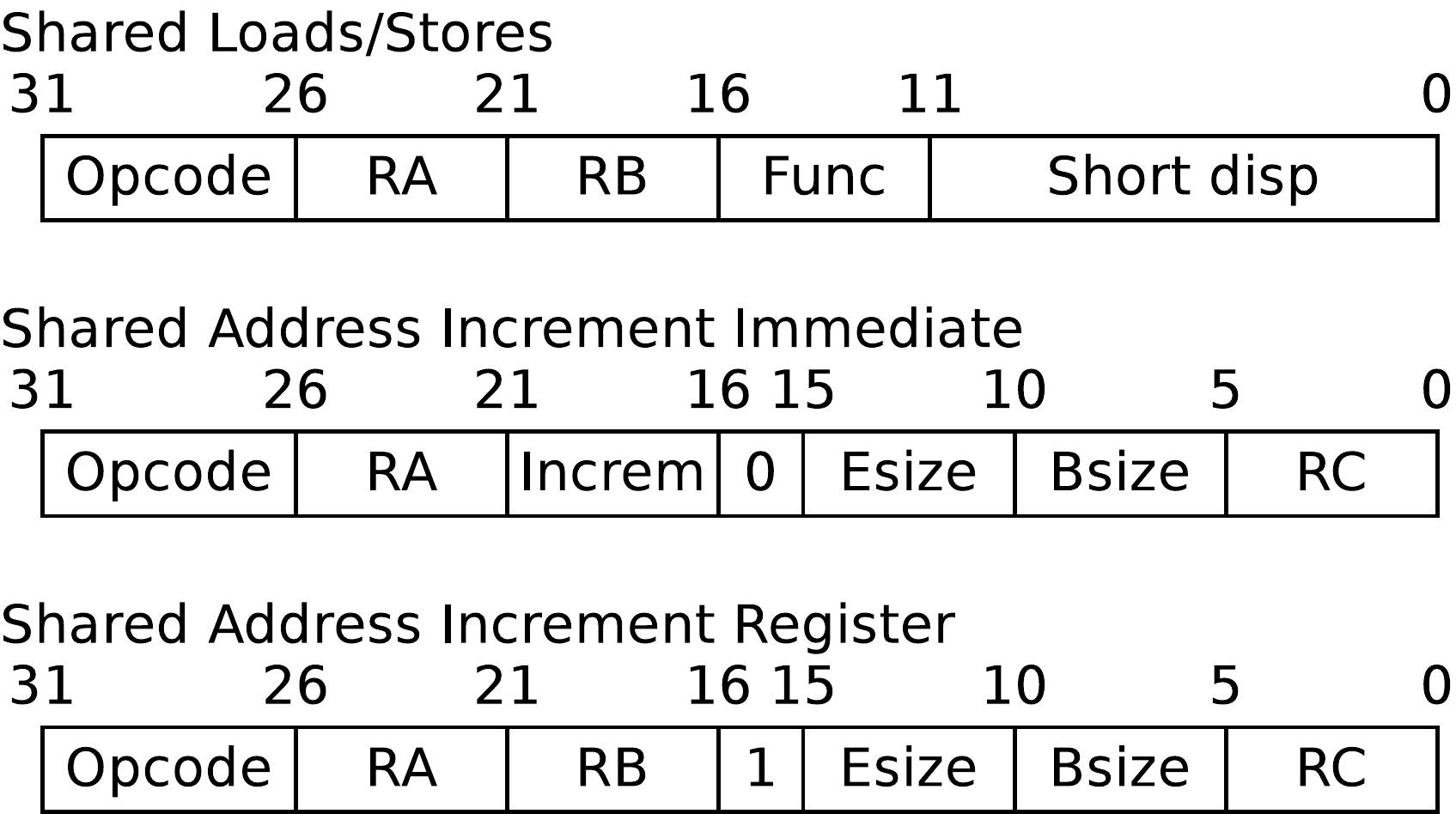}
    \caption{New Alpha Instructions Format}
    \label{fig:instruction_format}
\end{figure}

In order to simulate the extended instruction set, the Gem5 simulator
\cite{2011_nb_sigarch_gem5} was used.
Gem5 has multiple advantages for this work: it supports a great variety of
architectures, cache configurations and different CPU models allowing
trade-offs between speed of simulation and accuracy.
Gem5 has also been shown to be an accurate simulator
\cite{2012_ab_recosoc_gem5_accuracy}.

We selected the Alpha architecture as it supports full-system simulation up to
64 cores with GNU/Linux.
Gem5 simulates 64-bit Alpha 21264 processors with the BWX, CIX, FIX and MVI
extensions.
It provides 32 integer registers (R0-R31) and 32 floating point registers
(F0-F31).
We used the custom BigTsunami architecture in order to support up to 64 cores.
The Linux kernel version 2.6.27.62 patched for BigTsunami is used, programs
are compiled with the Berkeley UPC 2.14.2 compiler and cross compiled for Alpha
using GCC version 4.3.2.

The \textit{Classic} Gem5 memory model is used; each core is configured with a
32kb L1 code and data cache, and a shared L2 cache of 4MB.
The frequency is set at 2 Ghz.

The Alpha instruction set is extended with the instructions shown in
Table~\ref{tab:isa_extension}.
The instruction format is presented in Figure~\ref{fig:instruction_format}.
The integer registers (R0-R31) are also used to store the shared addresses.

For loads and stores, \texttt{RA} and \texttt{RB} represent the source and
destination registers.
\texttt{Opcode} is a free opcode from the Alpha instruction set.
\texttt{Func} defines which type of load or store is going to be performed.
\texttt{Short disp} is a displacement added to the resulting virtual address
after the shared pointer has been translated to the system virtual address.
\texttt{Short disp} is particularly useful in order to access different members
of a data structure.

For the shared addresses incrementation instructions, \texttt{RA} and
\texttt{RC} represents the source and destination registers.
\texttt{RB} is used in the register version of the instruction and specify the
increment register. Any increment value can be used when using a register.
\texttt{Esize}, \texttt{Bsize} and \texttt{Increm} are 5-bit encoded immediate
values for the element size, block size and the increment; they can represent
any 32 bits value in which only one bit is set (1, 2, 4, 8, ...).

In order to maintain the productivity advantage of UPC, the new instructions
need to be usable without any user intervention.
A prototype compiler was realized based on the Berkeley UPC source-to-source
compiler.
For that, the first step was to disable the phase-less pointer optimization; as
this optimization generates an incompatible shared pointer format and the
optimization is not required when the hardware support is present.
The second step was to replace the shared pointer operation amenable to
hardware with \texttt{asm()} statements making use of the new instructions.
This is not always possible; for example, block sizes that are not powers of
two.
In such cases, the normal software address incrementation is used.
Some simple optimizations are also performed, for example, shared address
incrementation with only two bits set in the increment are performed via two
immediates : to increment a pointer by 3, an incrementation by 1 is done,
followed by an incrementation by 2.
The C code generated by the source-to-source compiler is then compiled with
the GCC compiler.
The assembler was also modified in order to recognize the new instructions.

\subsection{Hardware Based Implementation}
\label{sec:fpga_based_implementation}

\begin{figure}[!tbp]
    \centering
    \includegraphics[width=\columnwidth]{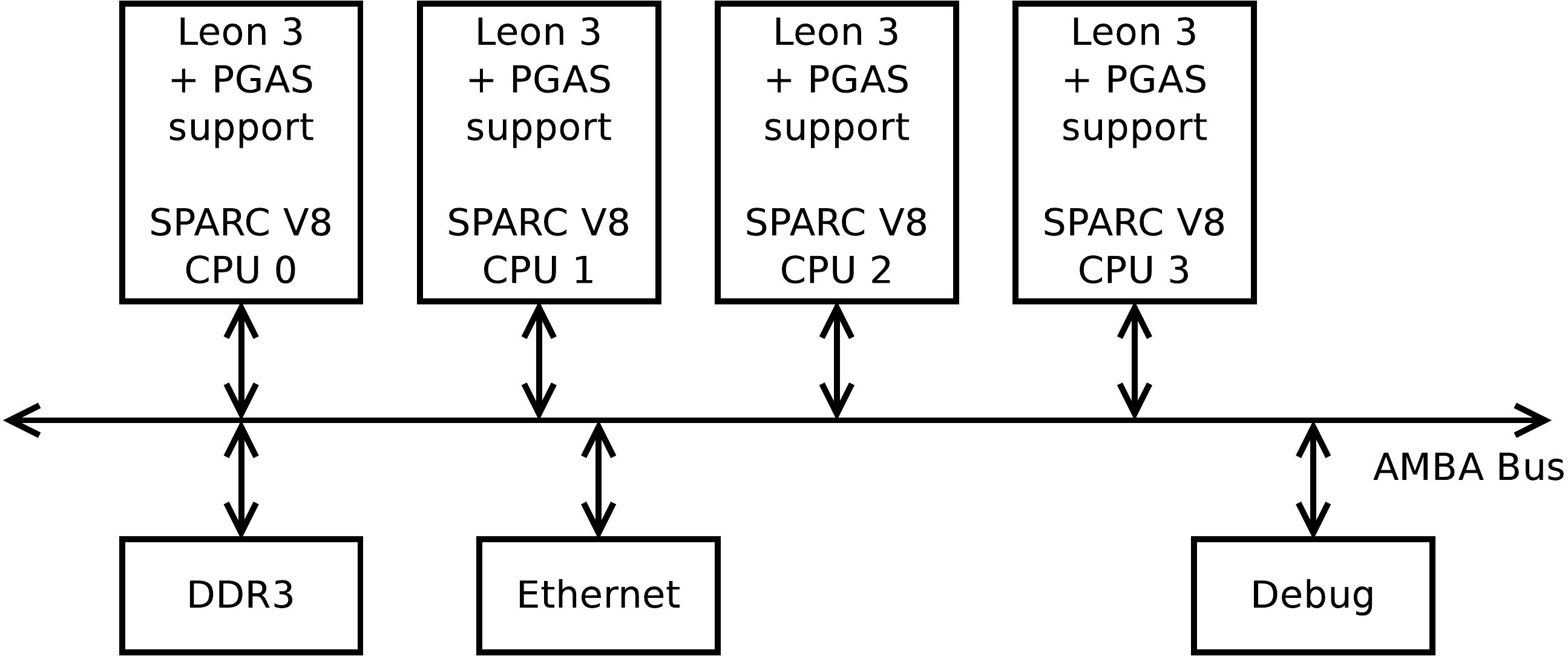}
    \caption{4 core Leon3 SMP with PGAS support}
    \label{fig:leon3smp}
\end{figure}

\begin{figure}[!tbp]
    \centering
    \includegraphics[width=\columnwidth]{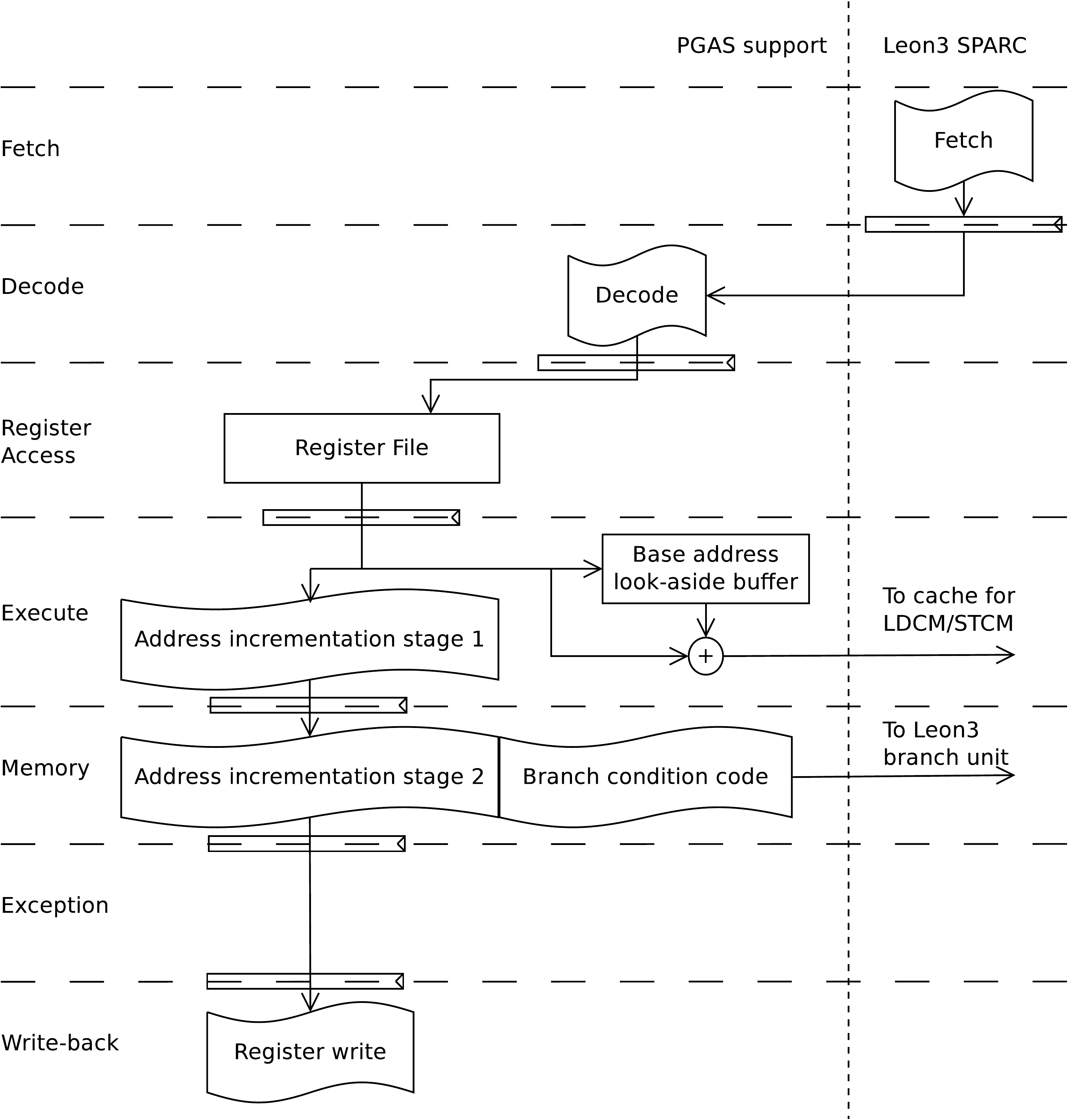}
    \caption{The 7-stage Leon3 pipeline extended with PGAS support}
    \label{fig:leon3pipeline}
\end{figure}

\begin{table}[!tbp]
    \caption{Leon3 configuration}
    \centering
    \begin{tabular}{|l|l|}
        \hline
                  & Configuration \\
        \hline
        Cores     & 4x SPARC cores (SMP) \\
        Features  & 2-cycle multiplier, branch prediction \\
        Cache     & Cache Coherent \\
        L1 I      & 2 Sets, 8 kB/set, 32 bytes/line, LRU \\
        L1 D      & 4 Sets, 4 kB/set, 16 bytes/line, LRU \\
        FPU       & Not implemented \\
        BUS       & AMBA AHB with fast snooping \\
        Memory    & Xilinx MIG-3.7 DDR3-800 \\
        Frequency & 75MHz \\
        OS        & GNU/Linux, Linux version 2.6.36 \\
        \hline
    \end{tabular}
\end{table}

\begin{table}[!tbp]
    \centering
    \caption{PGAS Hardware Support SPARC V8 ISA extension}
    \begin{tabular}{|l|l|}
        \hline
        \multicolumn{2}{|c|}{Coprocessor Load/Store} \\
        \hline
        LDC  & Load to Coproc. reg.\\
        & (32 bits)\\
        STC  & Store from Coproc. reg.\\
        & (32 bits)\\
        \hline
        \multicolumn{2}{|c|}{Shared Address Load/Store} \\
        \hline
        LDCM & Load Long (32 bits)\\
        STCM & Store Long (32 bits)\\
        \hline
        \multicolumn{2}{|c|}{Branch}\\
        \hline
        CB123 & Branch on locality\\
        \hline
        \multicolumn{2}{|c|}{Shared Address Incrementation} \\
        \hline
        SH\_ADD\_INC      & Immediate\\
        SH\_ADD\_INC\_REG & Register\\
        \hline
    \end{tabular}
    \label{tab:isa_extension_sparc}
\end{table}

In order to evaluate the implementability of the proposed solution on real
hardware, we realized a prototype on FPGA.
For that, a softcore processor (Leon3) was extended with hardware support for
PGAS.

The Leon3 softcore processor implements the 32-bit SPARC V8 architecture with a
7-stage pipeline.
It has the advantage of supporting cache coherent SMP systems allowing it to the full GNU/Linux
operating system.
The VHDL source code of the base Leon3 is available under a GNU Public License
(GPL), allowing for modification.
The Leon softcore processor has already been used to study various specific
hardware support possibilities \cite{2006_gdm_mixdes_leon2_mpeg_encoding,
2010_md_ddecs_leon3_instruction_multi_threading}.

We extended the processor with hardware support for PGAS shared addresses by
using the reserved SPARC V8 instructions for a coprocessor.
More information about extending the Leon3 softcore processor via the
coprocessor interface can be found in
\cite{2011_os_reconfig_architecture_reconfigurable_multi_core}.

The coprocessor instructions are fully integrated with the main processor
pipeline : the instructions are fetched by the main pipeline and the
coprocessor instruction execution is synchronized with the main pipeline.
A register file is introduced for storing the shared address pointers as those
are 64 bits; it is similar in all points with the register file used for
floating point support in Leon3.
It enables reading two 64-bits values per clock cycle and writing one 64-bits
value. The extra register would not be needed for 64 bits architectures as
it is the case for our GEM5 Alpha implementation.

Figure~\ref{fig:leon3pipeline} presents the Leon3 pipeline extended with
hardware support for shared pointers.
The address incrementation is fully pipelined over two stages, allowing to
perform one address translation per clock cycle.
It also generates a coprocessor condition code based on the locality of the
incremented address.
Four condition codes are possible: 0: local (the pointed data is own by the
current thread), 1: located on the same memory controller, 2: accessible by the
load/store from shared instructions, 3: located on an other node.
The Coprocessor Branch (CB) instruction allows to branch based on any
combination of the condition code.
Loads and stores from shared addresses (LDCM, STCM) are performed as fast as
the normal SPARC load and store instructions.

The design was implemented on a Virtex-6 FPGA ML605 Evaluation board.
The ML605 is based on a Virtex 6 XC6VLX240T-1FFG1156 FPGA.
The logic synthesis, place and route for the design was performed using Xilinx
ISE Release 13.4. The final design runs at a frequency of 75 MHz.

\section{Results}
\label{sec:results}

In this section, the results from the simulation and the hardware
implementation are presented.

\subsection{Simulation Results}

\begin{figure*}[!tbp]
    \centering
    \subfloat[Performance normalized to the code without manual optimization]{
        \includegraphics[width=\graphwidth]{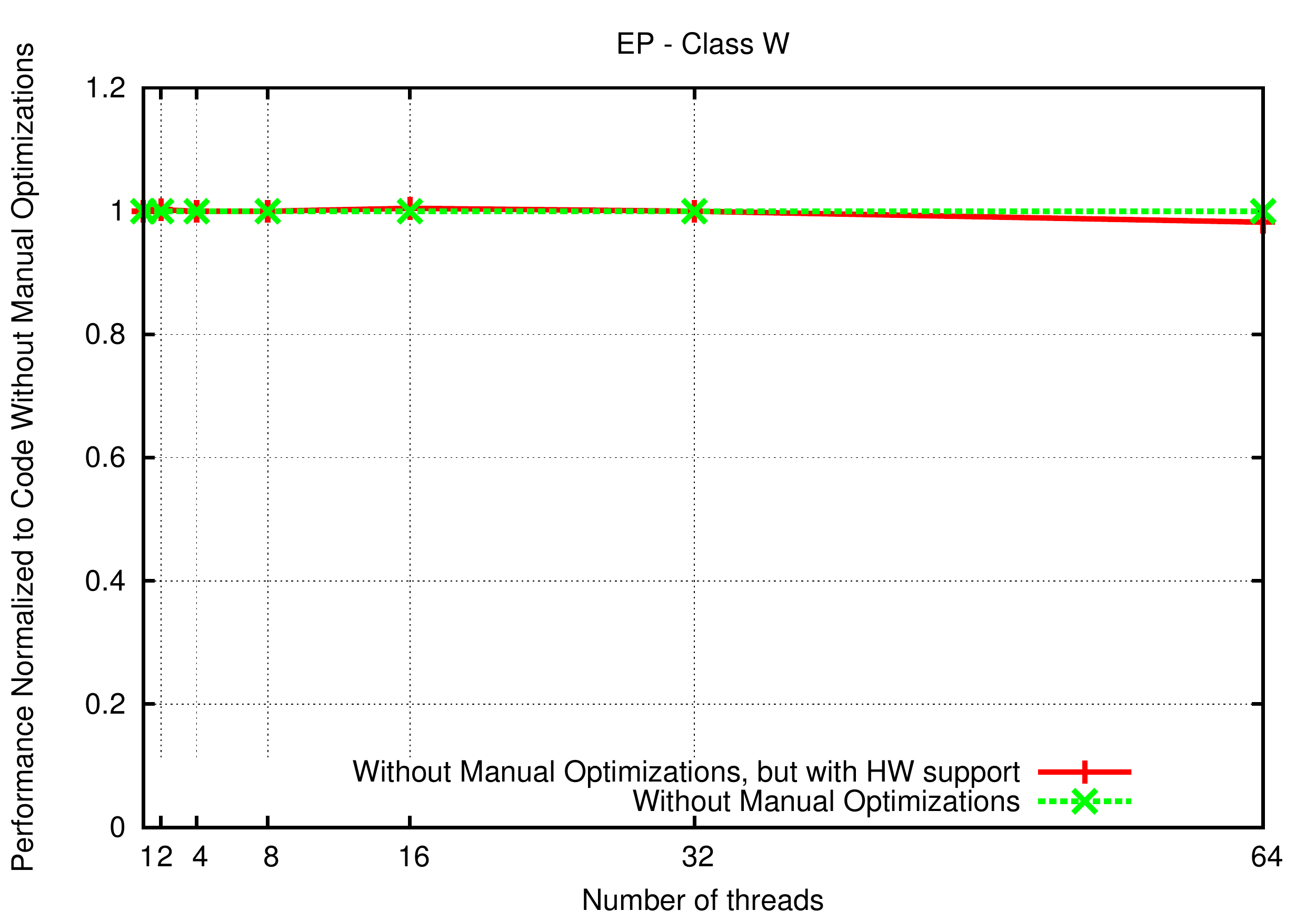}
    }
    \subfloat[Execution Time]{
        \includegraphics[width=\graphwidth]{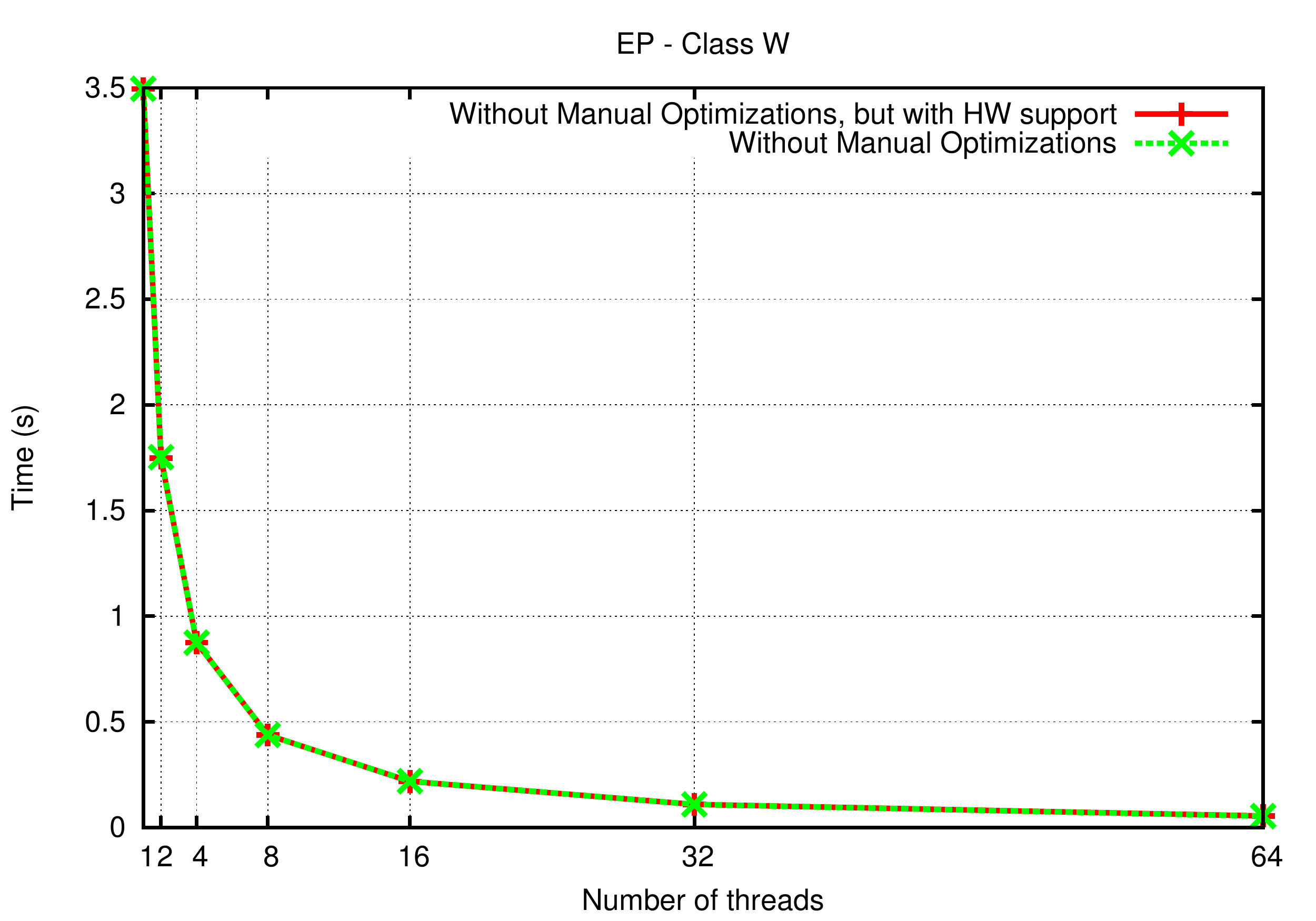}
    }
    \caption{Gem5 atomic model: NAS Parallel Benchmark - EP class W}
    \label{fig:results_ep_atomic}
\end{figure*}

\begin{figure*}[!tbp]
    \centering
    \subfloat[Performance normalized to the code without manual optimization]{
        \includegraphics[width=\graphwidth]{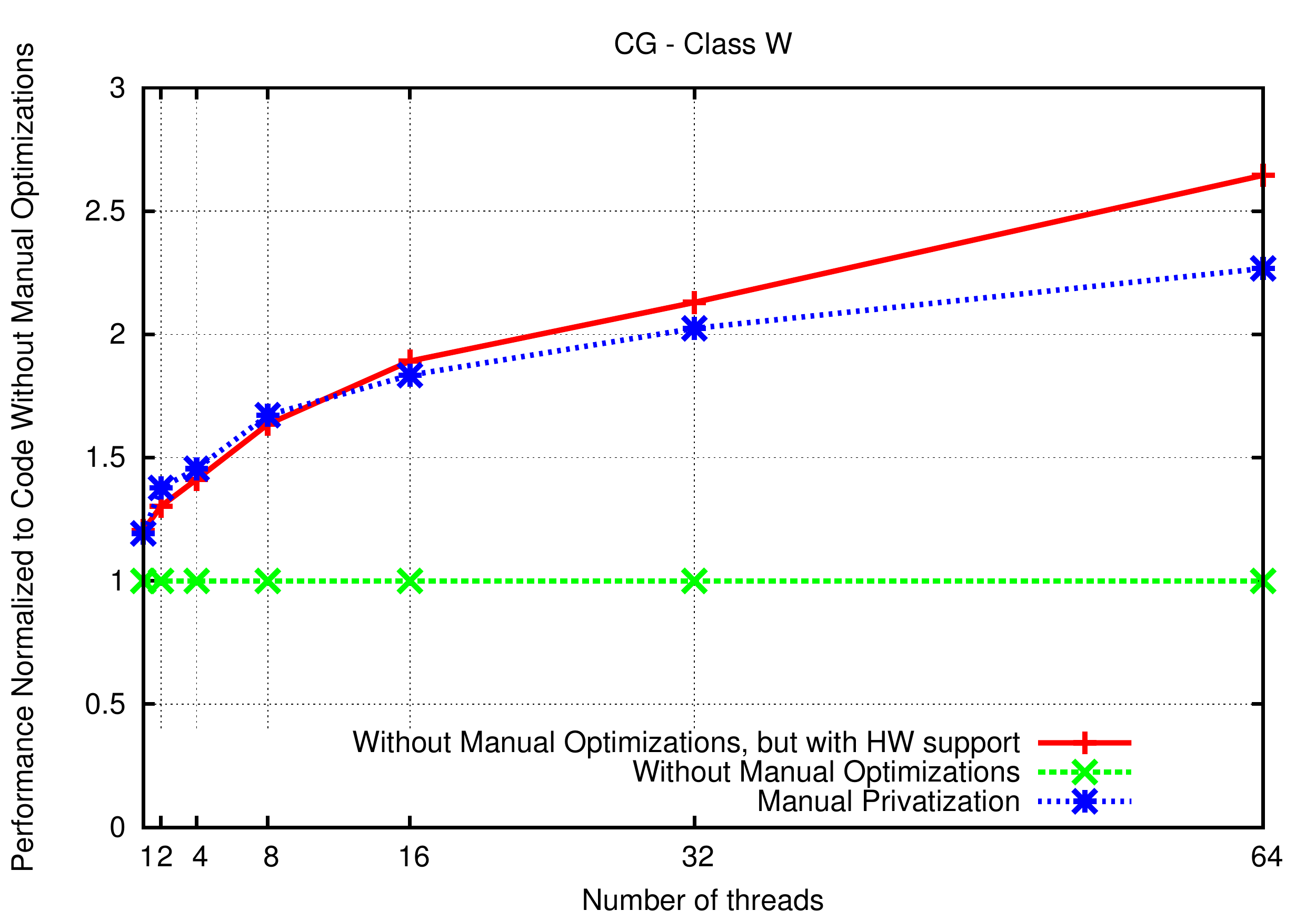}
    }
    \subfloat[Execution Time]{
        \includegraphics[width=\graphwidth]{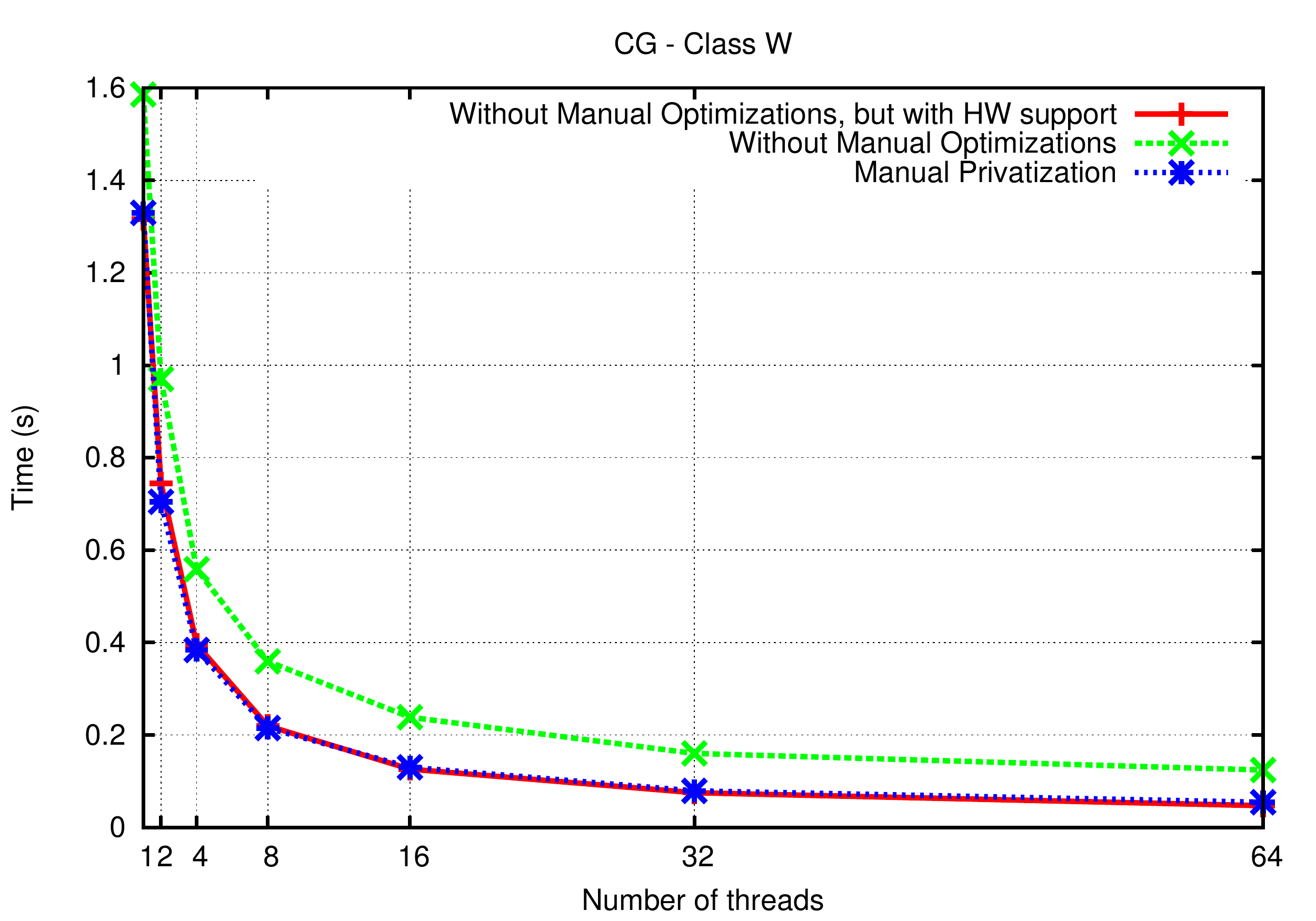}
    }
    \caption{Gem5 atomic model: NAS Parallel Benchmark - CG class W}
    \label{fig:results_cg_atomic}
\end{figure*}

\begin{figure*}[!tbp]
    \centering
    \subfloat[Performance normalized to the code without manual optimization]{
        \includegraphics[width=\graphwidth]{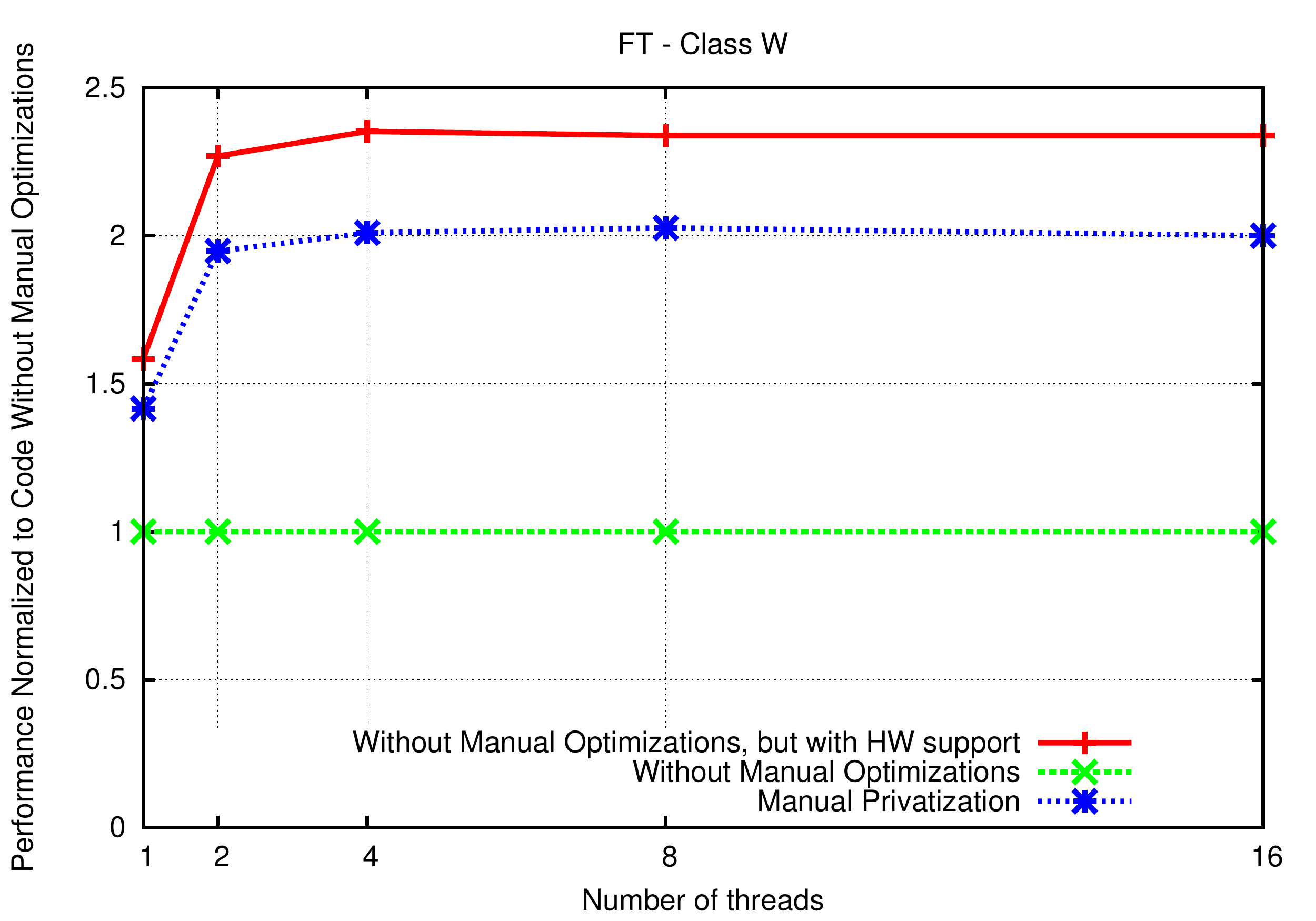}
    }
    \subfloat[Execution Time]{
        \includegraphics[width=\graphwidth]{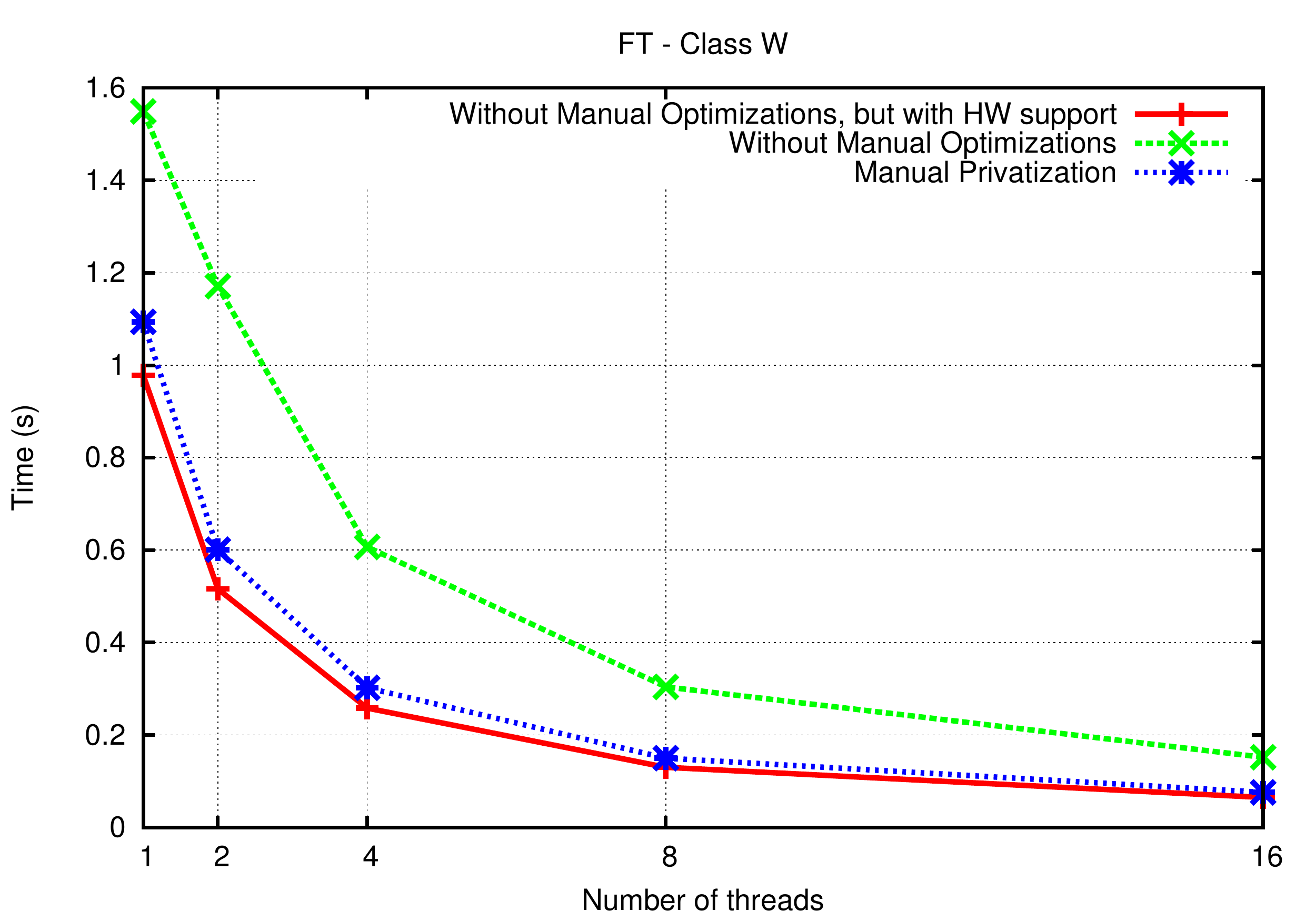}
    }
    \caption{Gem5 atomic model: NAS Parallel Benchmark - FT class W}
    \label{fig:results_ft_atomic}
\end{figure*}

\begin{figure*}[!tbp]
    \centering
    \subfloat[Performance normalized to the code without manual optimization]{
        \includegraphics[width=\graphwidth]{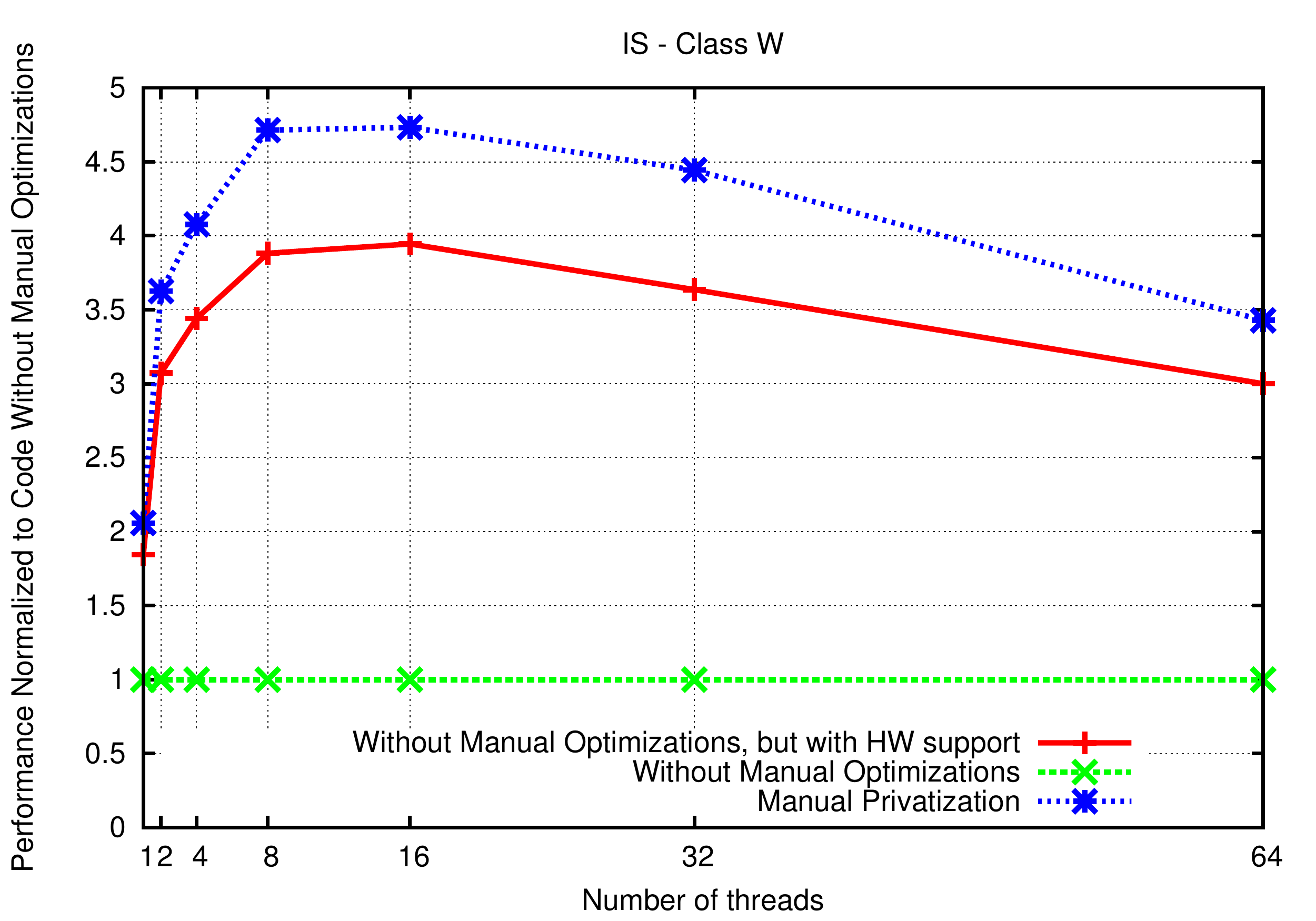}
    }
    \subfloat[Execution Time]{
        \includegraphics[width=\graphwidth]{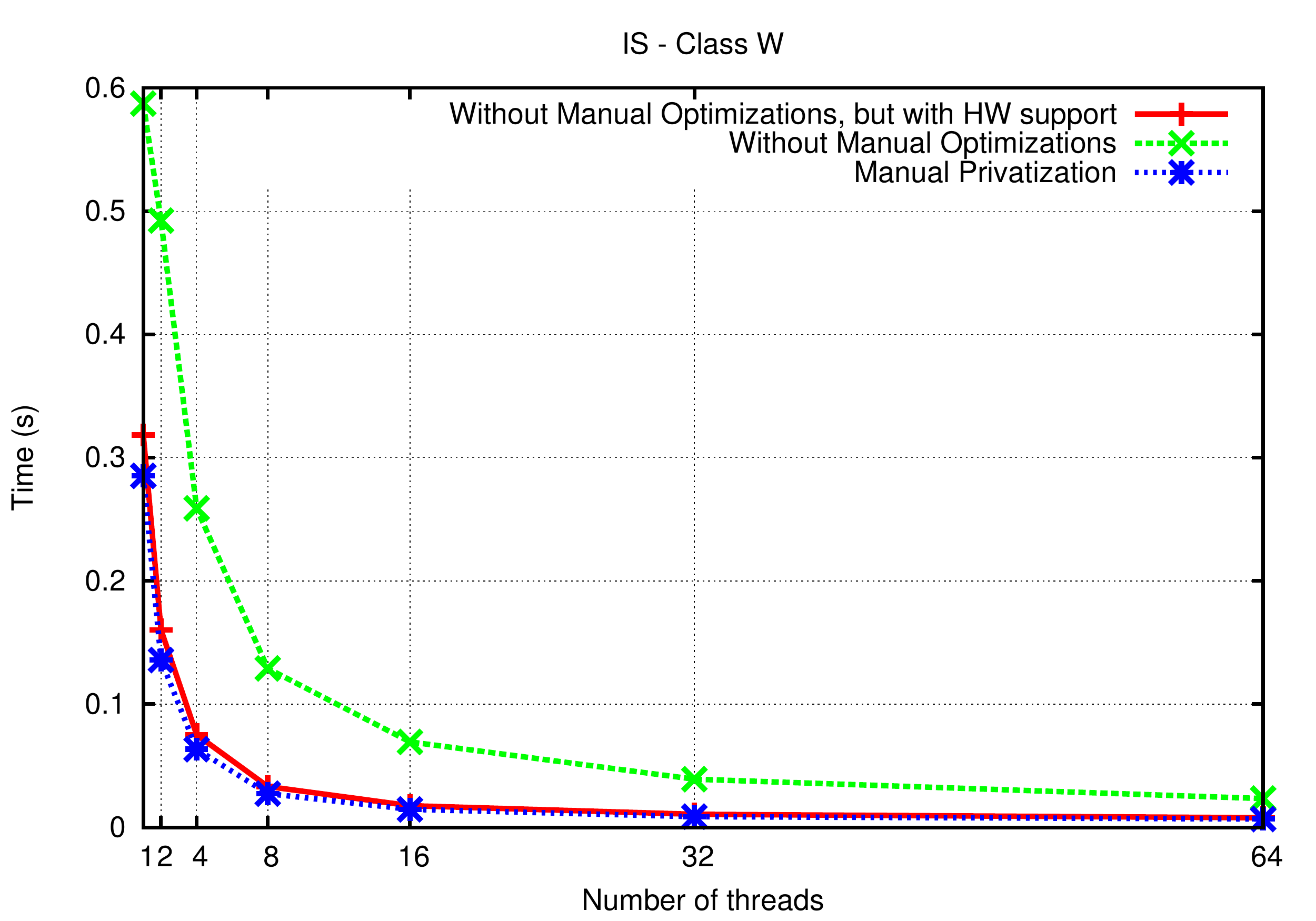}
    }
    \caption{Gem5 atomic model: NAS Parallel Benchmark - IS class W}
    \label{fig:results_is_atomic}
\end{figure*}

\begin{figure*}[!tbp]
    \centering
    \subfloat[Performance normalized to the code without manual optimization]{
        \includegraphics[width=\graphwidth]{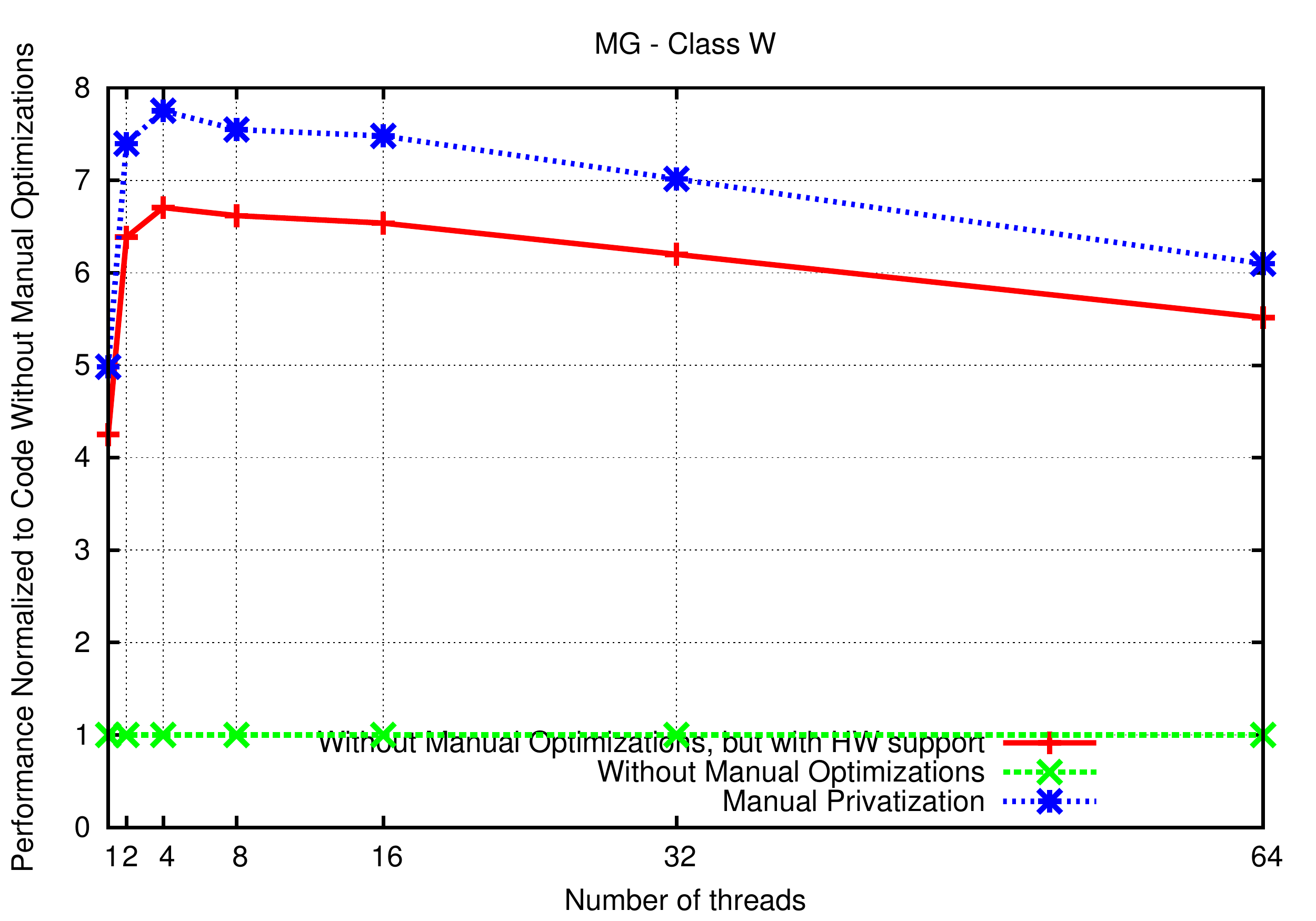}
    }
    \subfloat[Execution Time]{
        \includegraphics[width=\graphwidth]{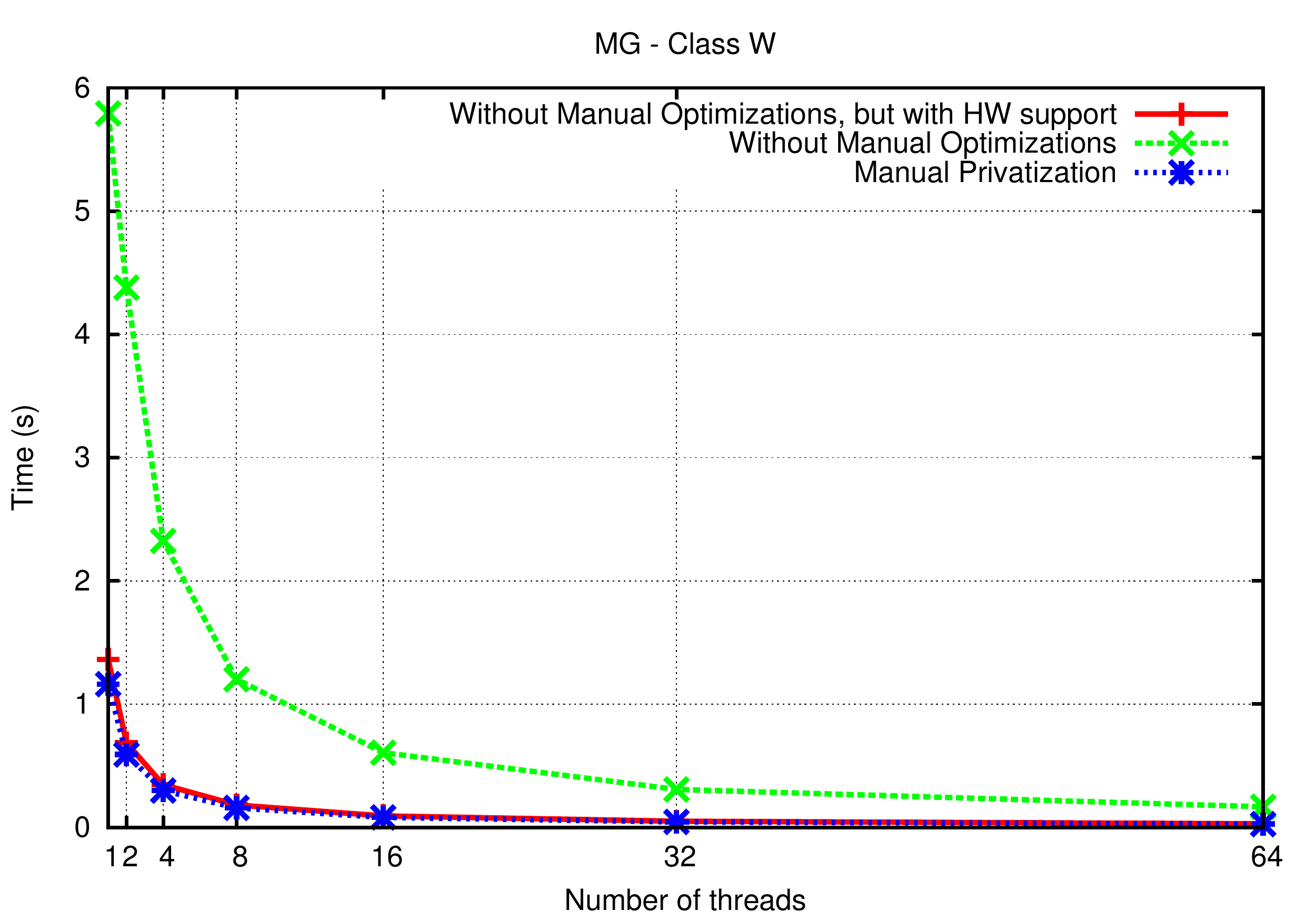}
    }
    \caption{Gem5 atomic model: NAS Parallel Benchmark - MG class W}
    \label{fig:results_mg_atomic}
\end{figure*}

\begin{figure*}[!tbp]
    \centering
    \subfloat[Timing - Improvement]{
        \includegraphics[width=\halfgraphwidth]{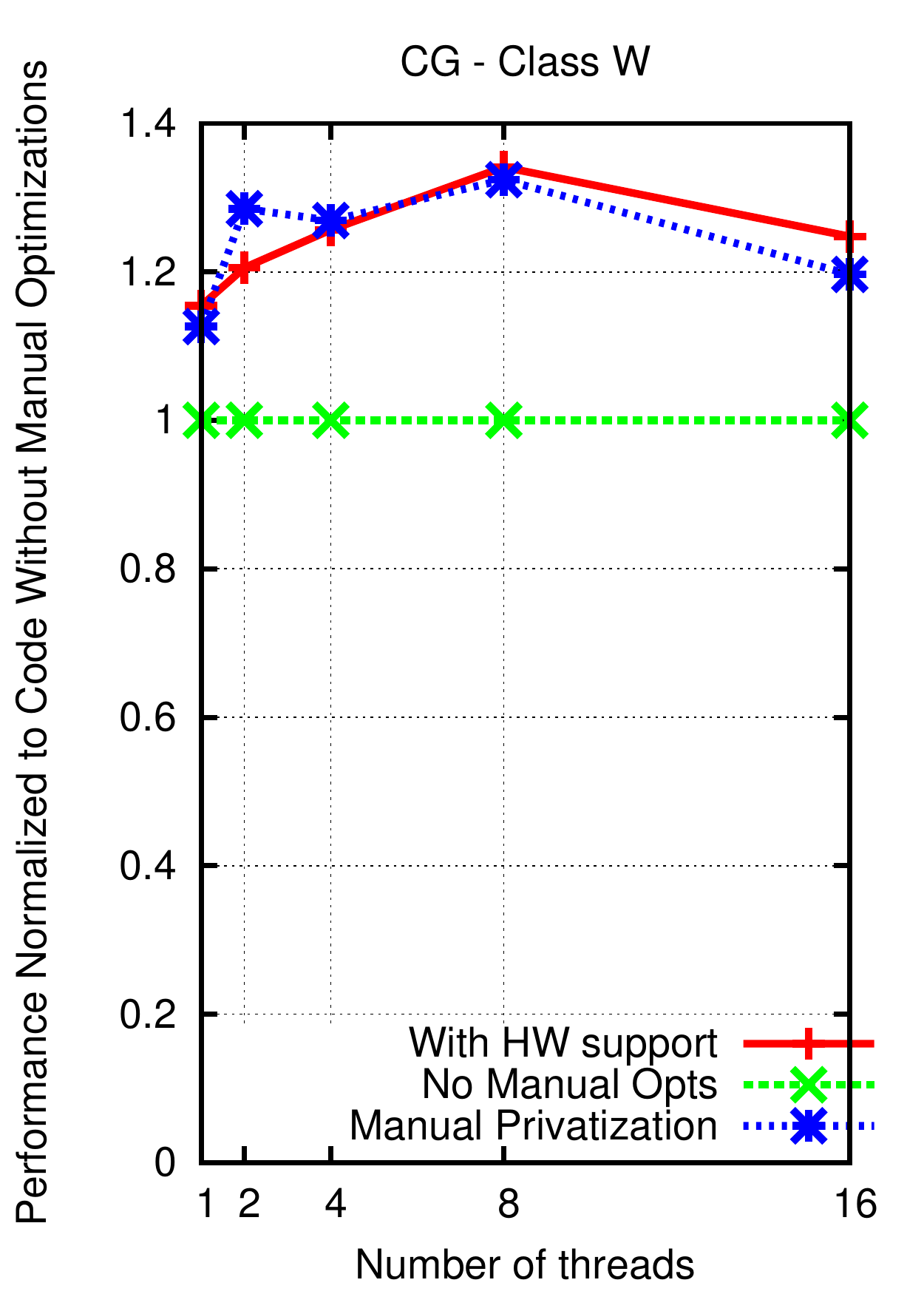}
    }
    \subfloat[Timing - Execution Time]{
        \includegraphics[width=\halfgraphwidth]{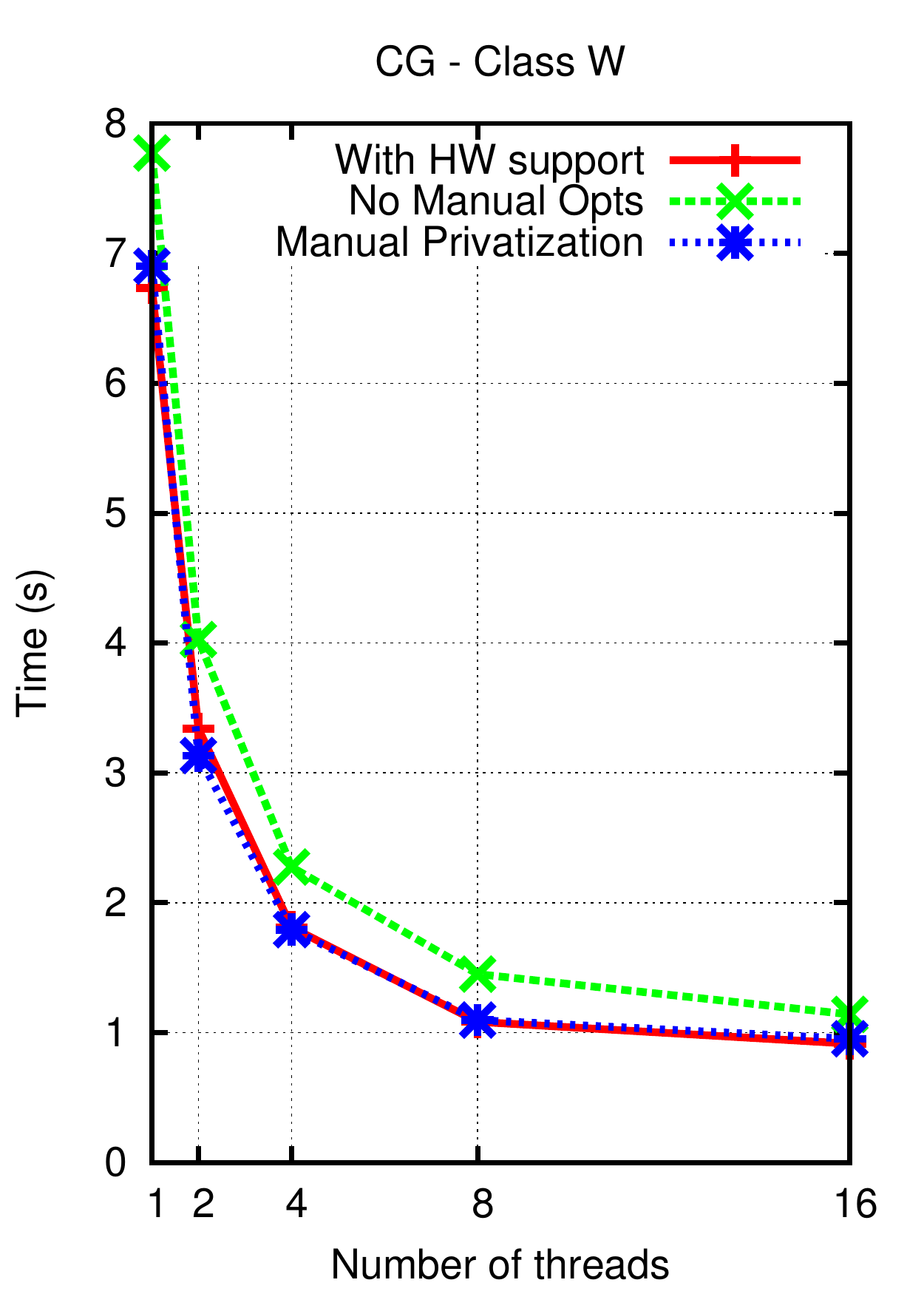}
    }
    \subfloat[Detailed - Improvement]{
        \includegraphics[width=\halfgraphwidth]{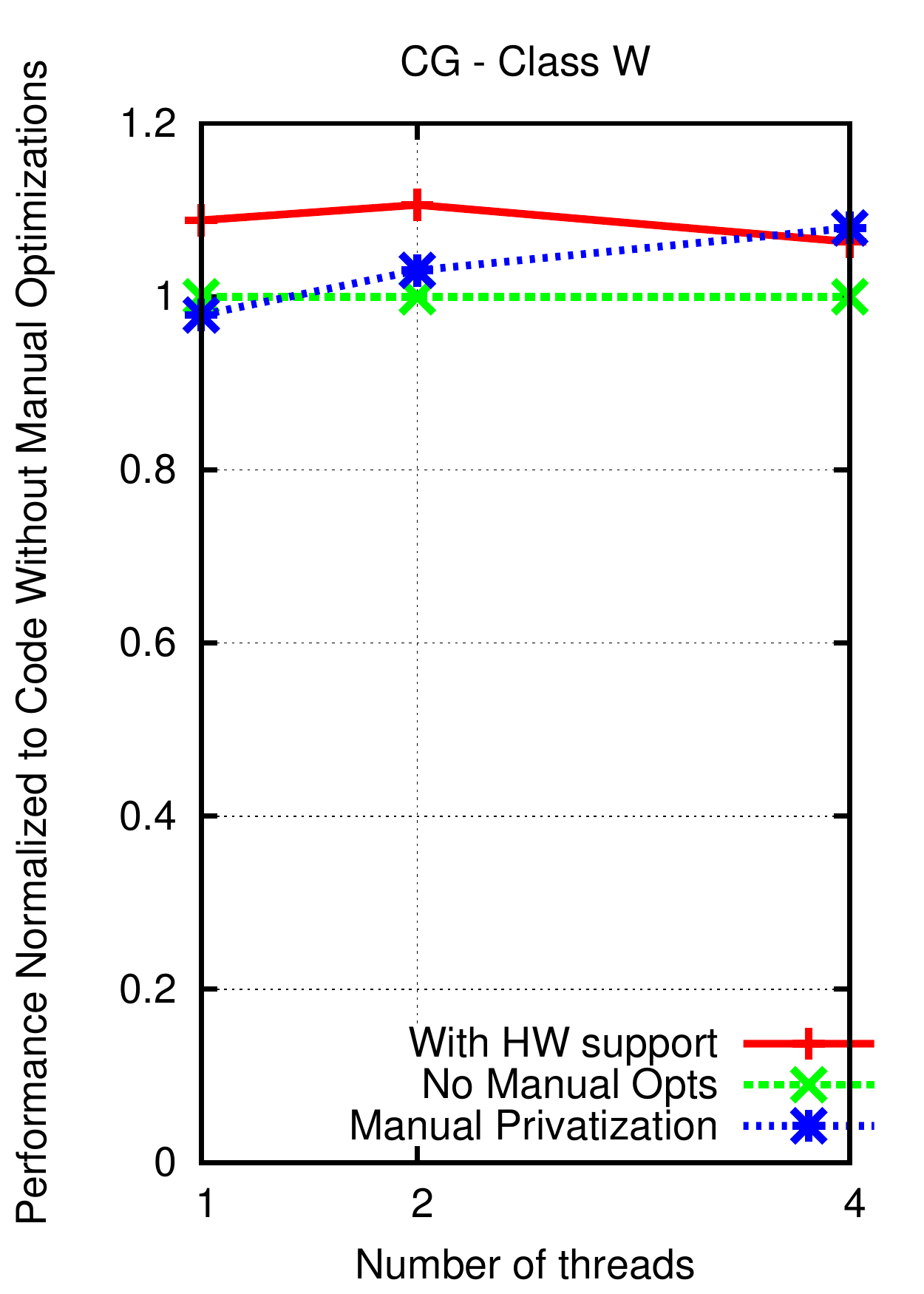}
    }
    \subfloat[Detailed - Execution Time]{
        \includegraphics[width=\halfgraphwidth]{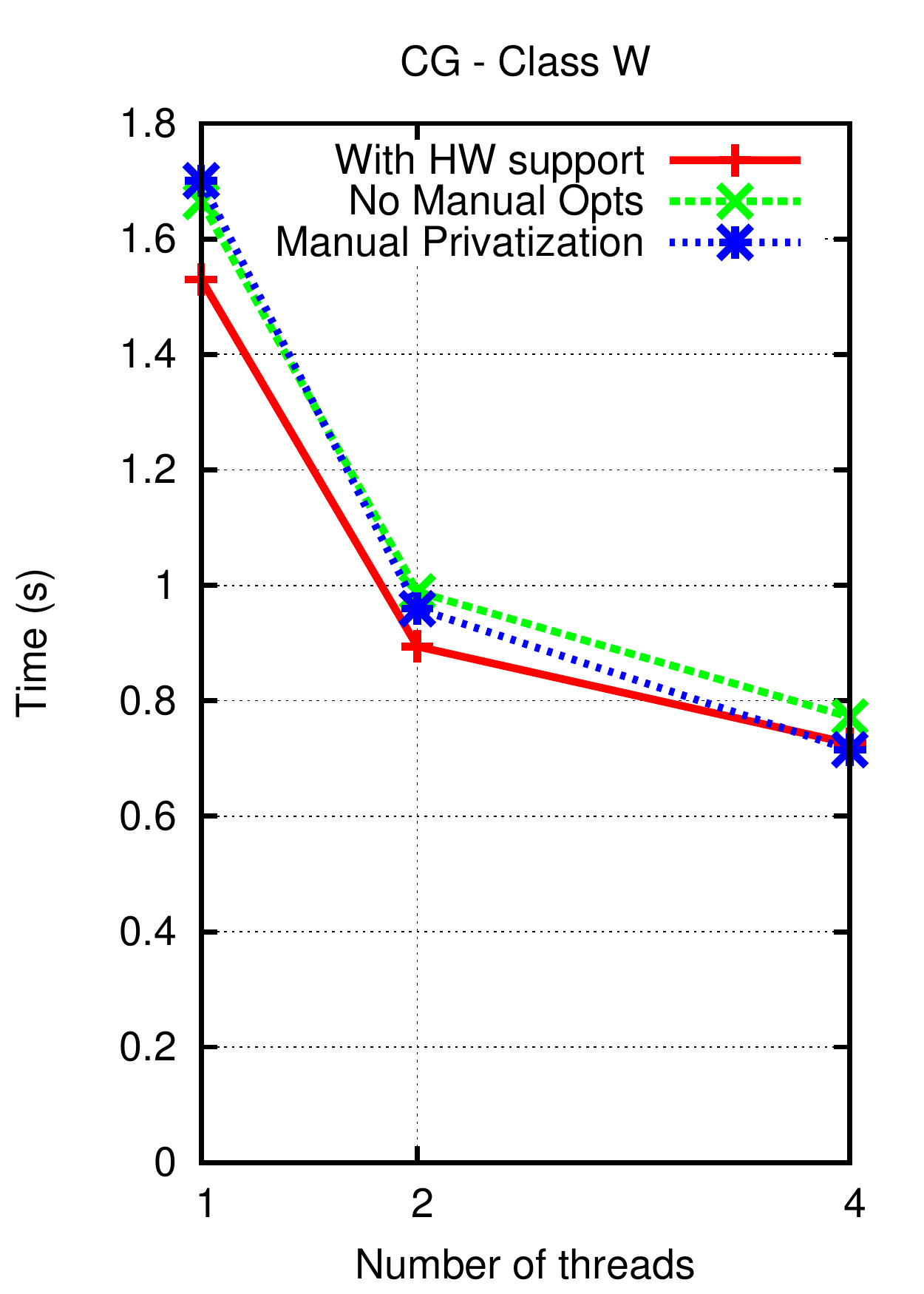}
    }
    \caption{Gem5 : NAS Parallel Benchmark - CG class W}
    \label{fig:results_cg}
\end{figure*}

\begin{figure*}[!tbp]
    \centering
    \subfloat[Timing - Improvement]{
        \includegraphics[width=\halfgraphwidth]{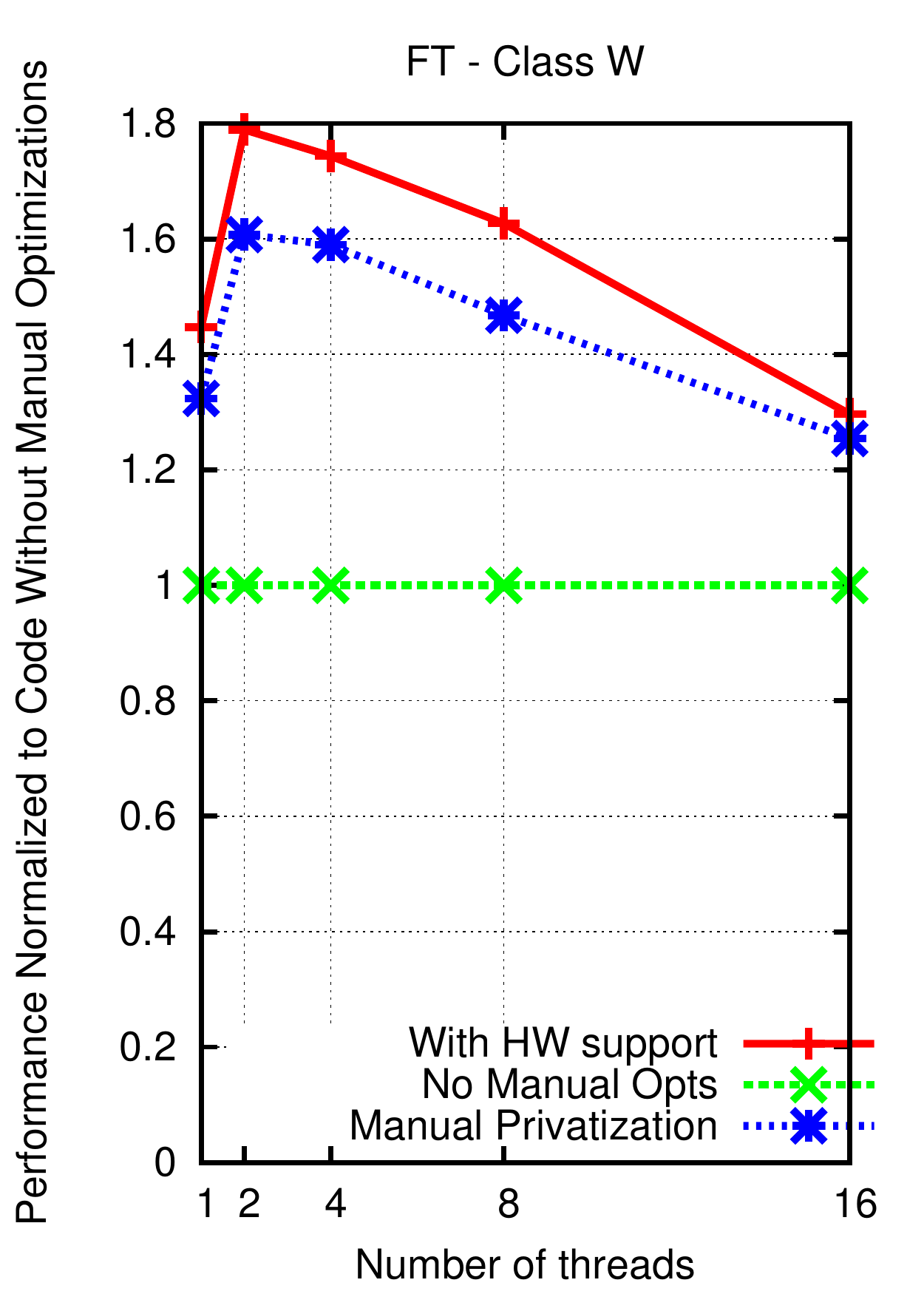}
    }
    \subfloat[Timing - Execution Time]{
        \includegraphics[width=\halfgraphwidth]{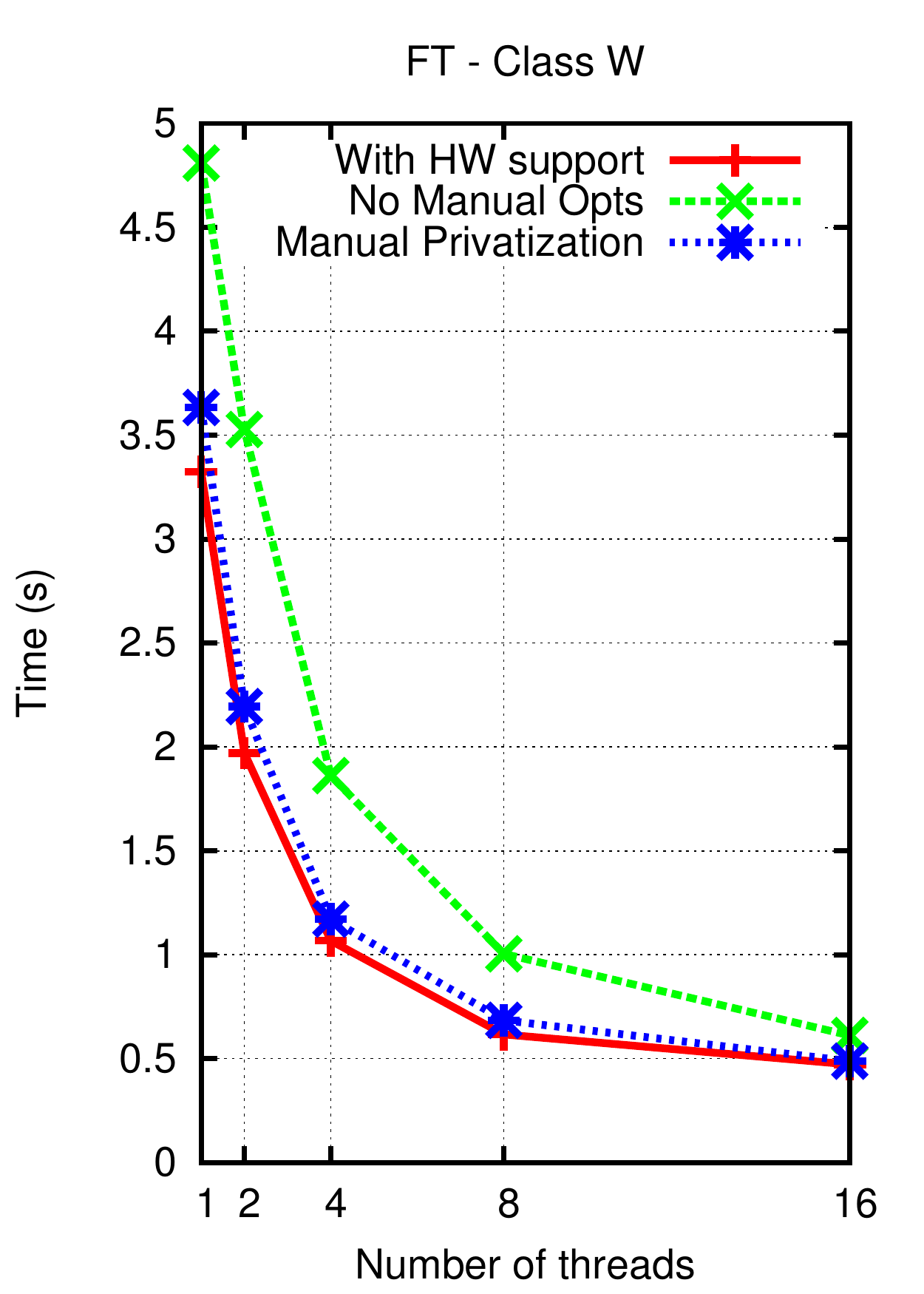}
    }
    \subfloat[Detailed - Improvement]{
        \includegraphics[width=\halfgraphwidth]{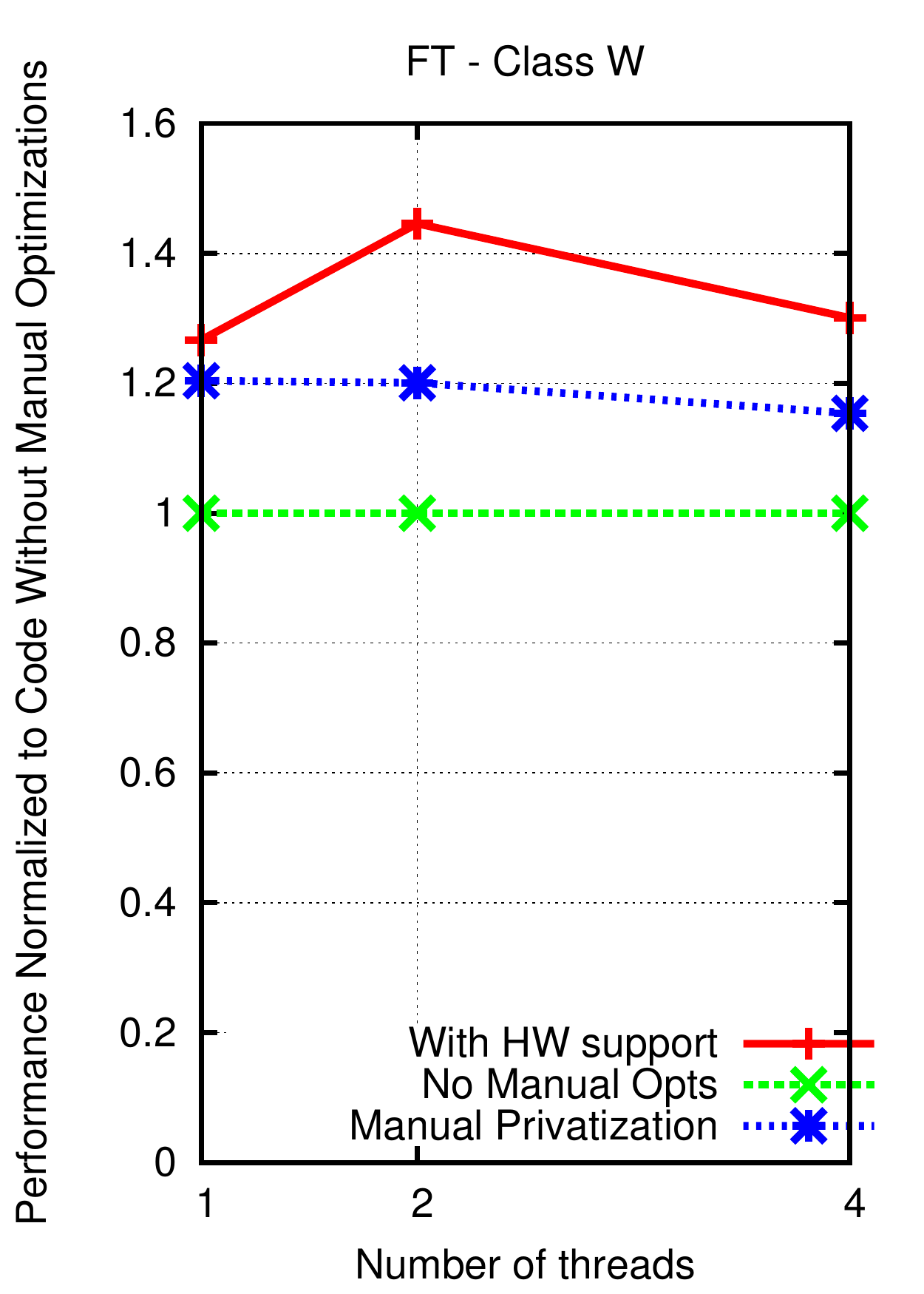}
    }
    \subfloat[Execution Time]{
        \includegraphics[width=\halfgraphwidth]{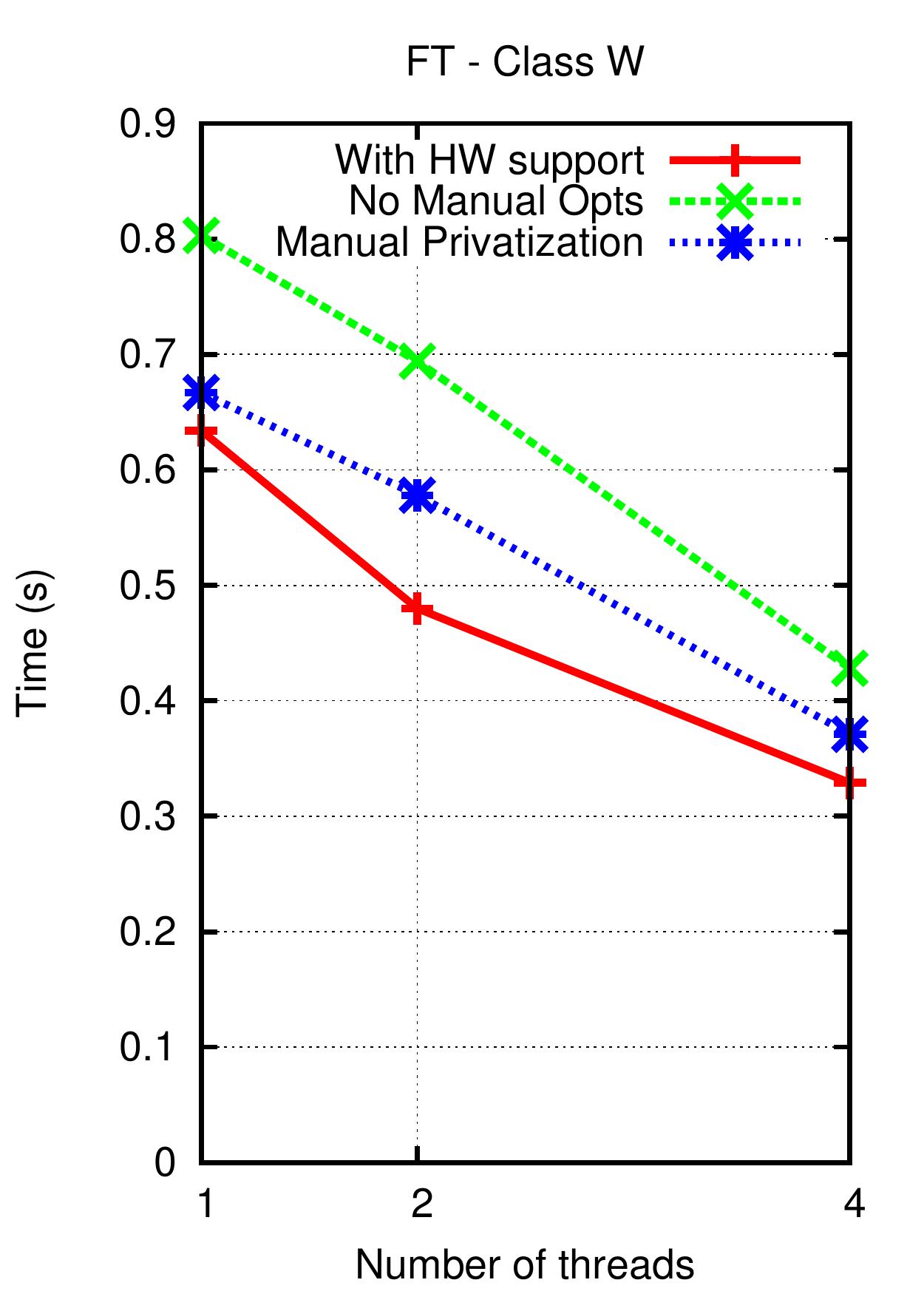}
    }
    \caption{Gem5 : NAS Parallel Benchmark - FT class W}
    \label{fig:results_ft}
\end{figure*}

\begin{figure*}[!tbp]
    \centering
    \subfloat[Timing - Improvement]{
        \includegraphics[width=\halfgraphwidth]{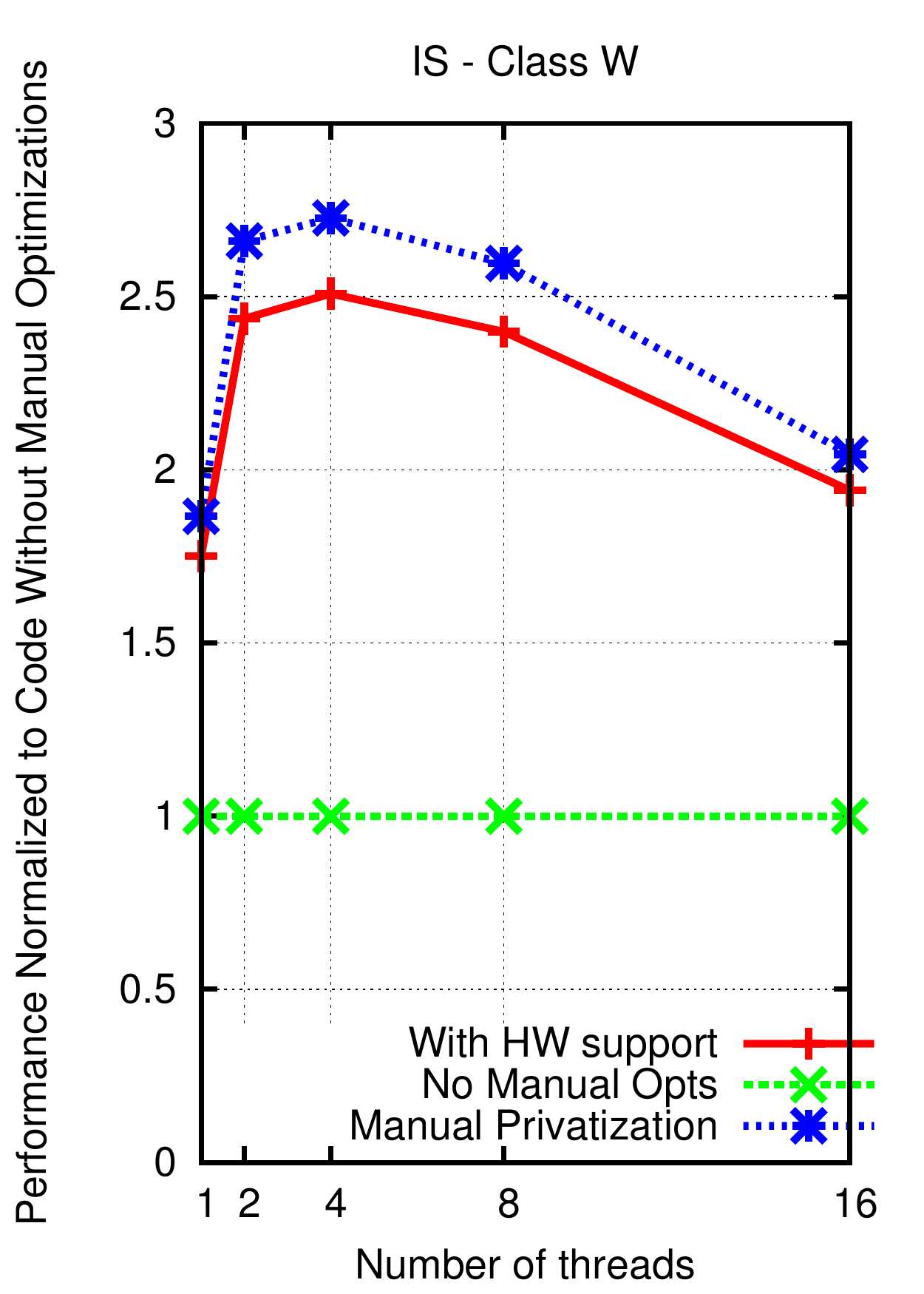}
    }
    \subfloat[Timing - Execution Time]{
        \includegraphics[width=\halfgraphwidth]{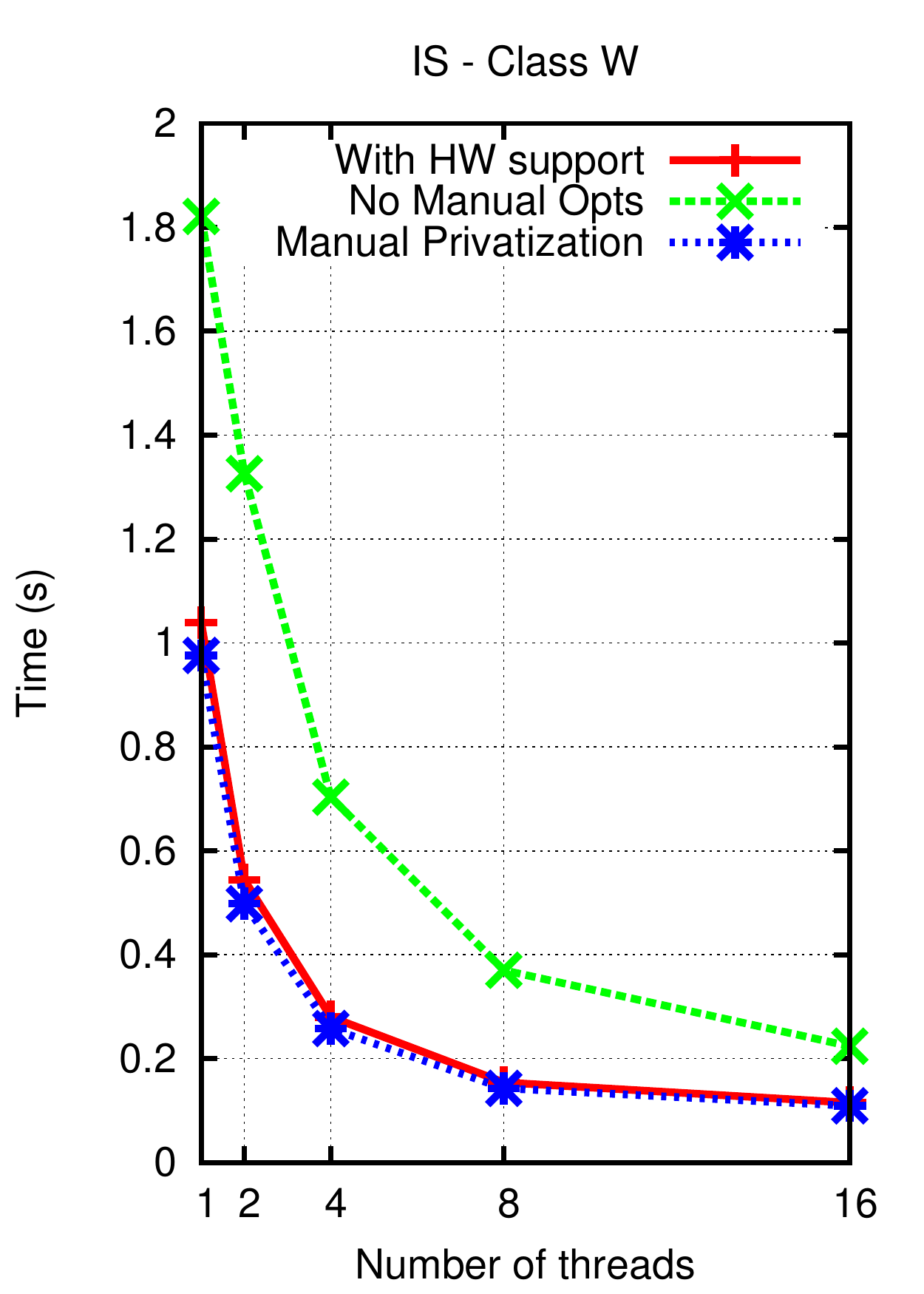}
    }
    \subfloat[Detailed - Improvement]{
        \includegraphics[width=\halfgraphwidth]{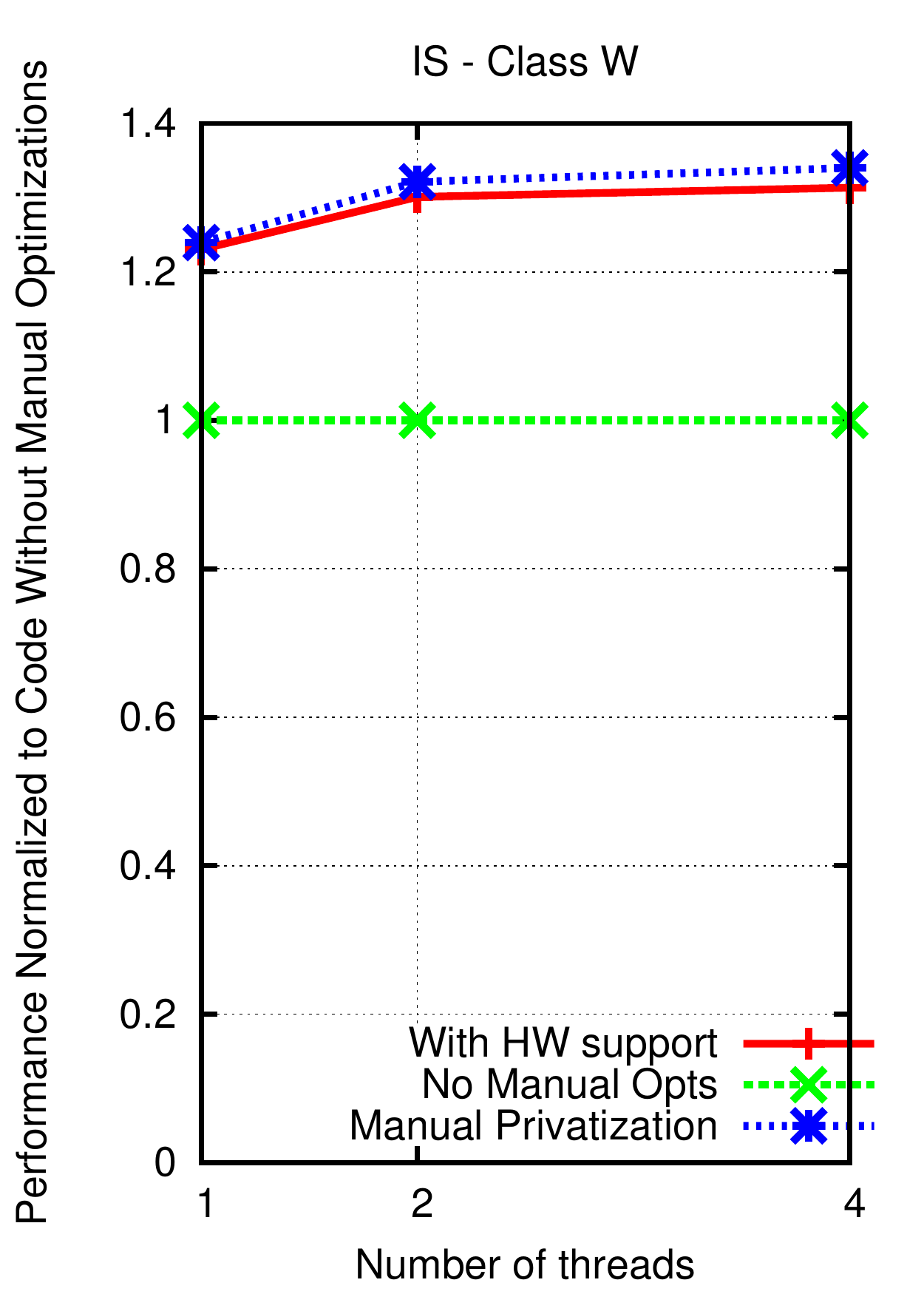}
    }
    \subfloat[Detailed - Execution Time]{
        \includegraphics[width=\halfgraphwidth]{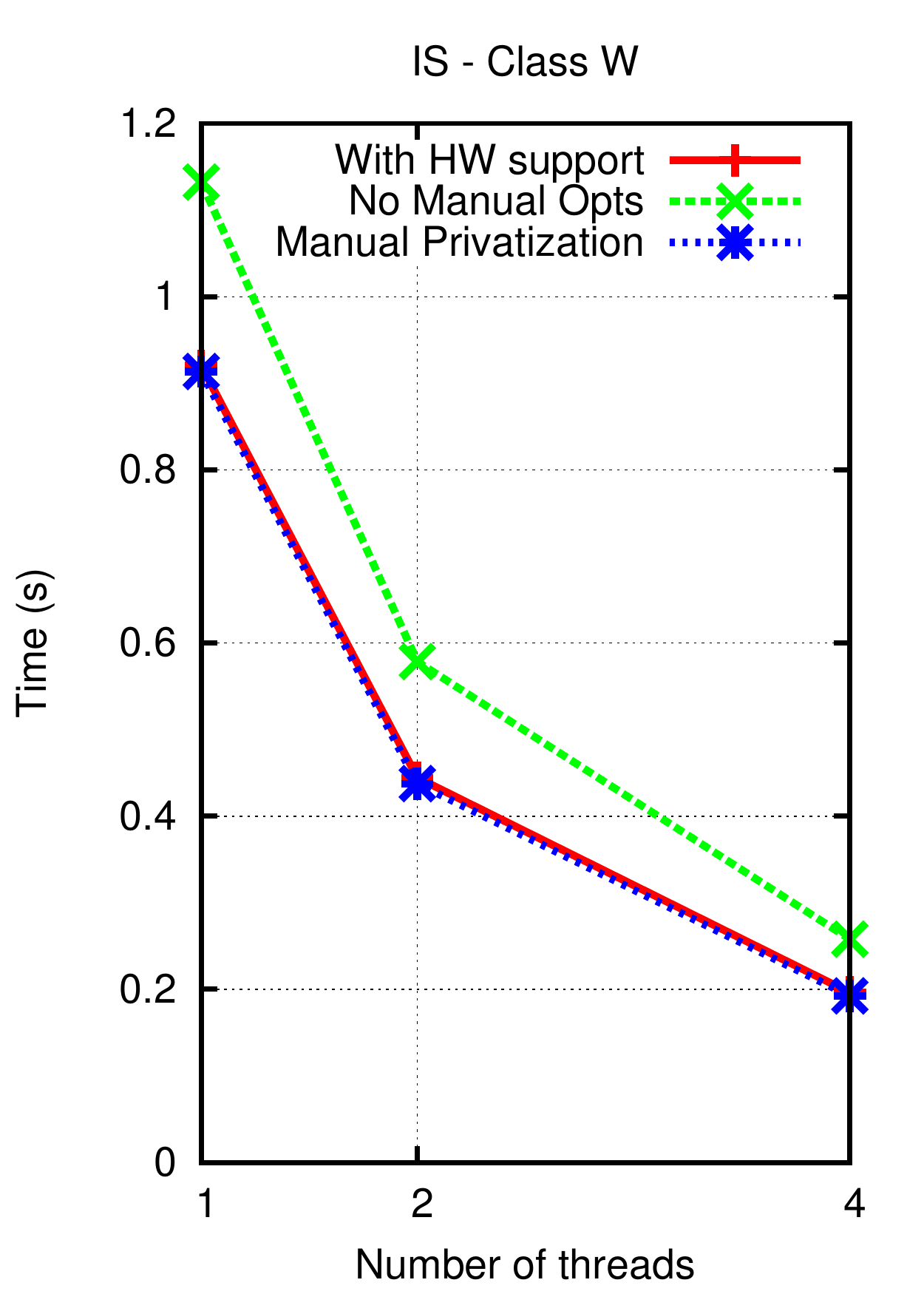}
    }
    \caption{Gem5 : NAS Parallel Benchmark - IS class W}
    \label{fig:results_is}
\end{figure*}

\begin{figure*}[!tbp]
    \centering
    \subfloat[Timing - Improvement]{
        \includegraphics[width=\halfgraphwidth]{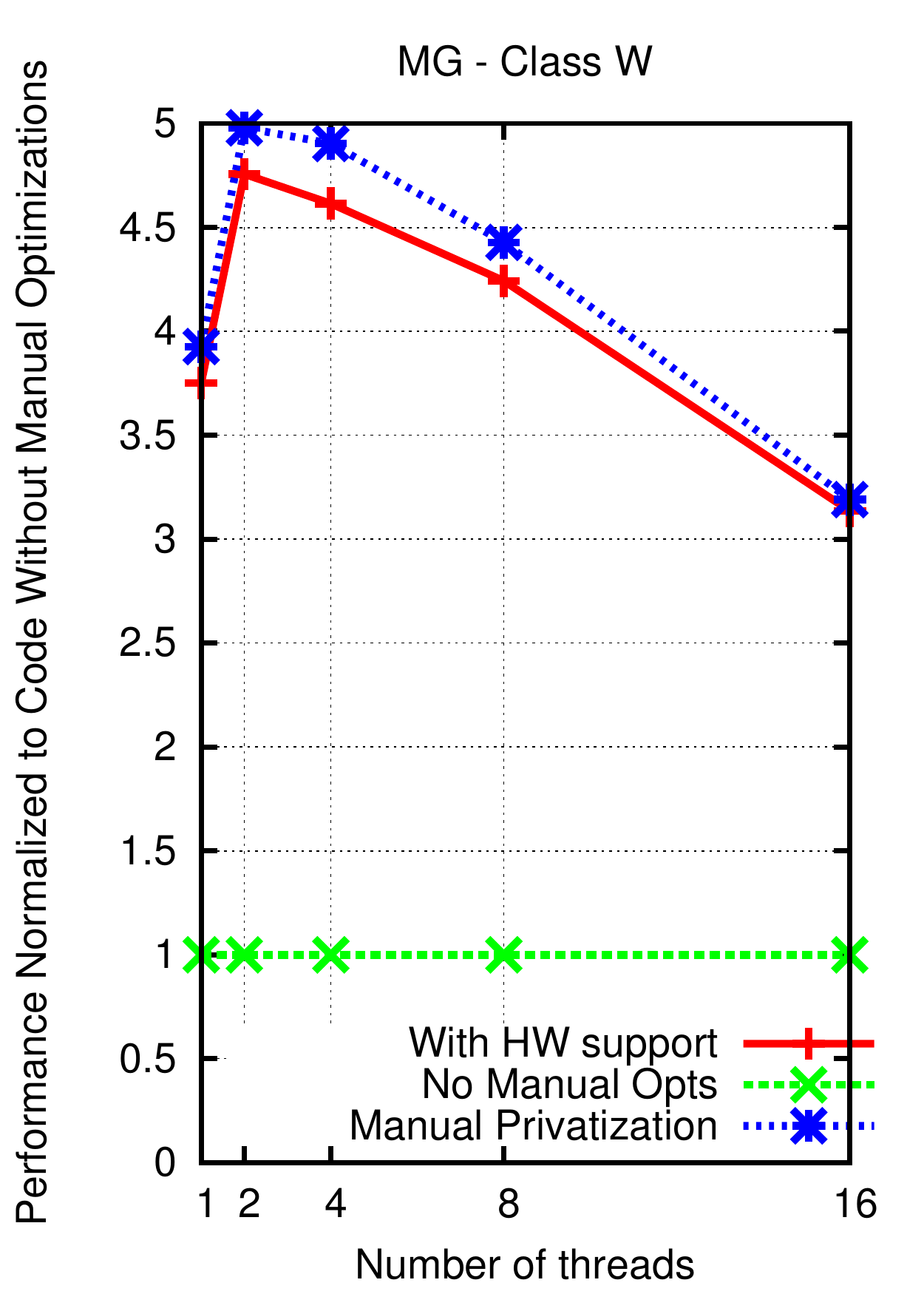}
    }
    \subfloat[Timing - Execution Time]{
        \includegraphics[width=\halfgraphwidth]{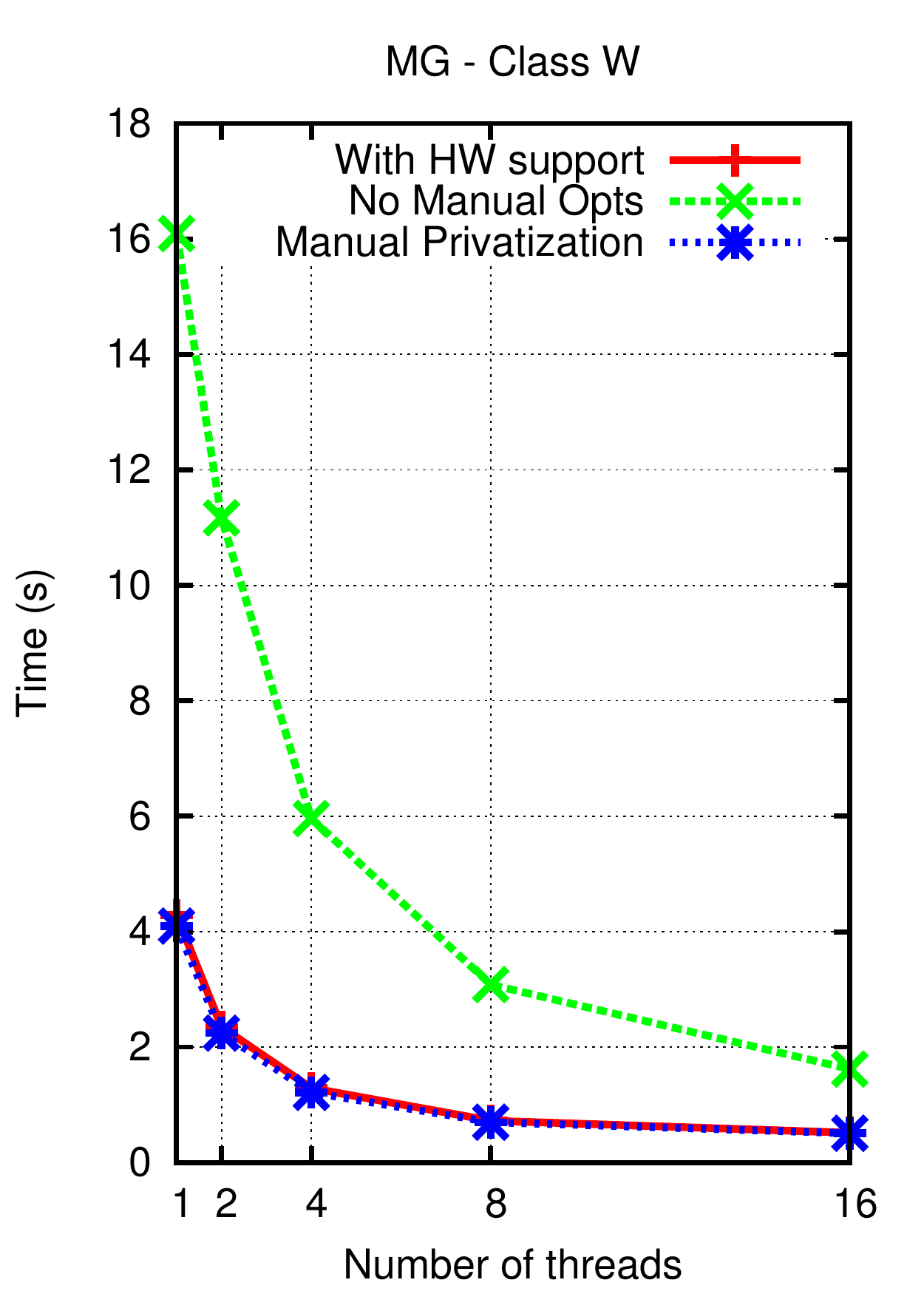}
    }
    \subfloat[Detailed - Improvement]{
        \includegraphics[width=\halfgraphwidth]{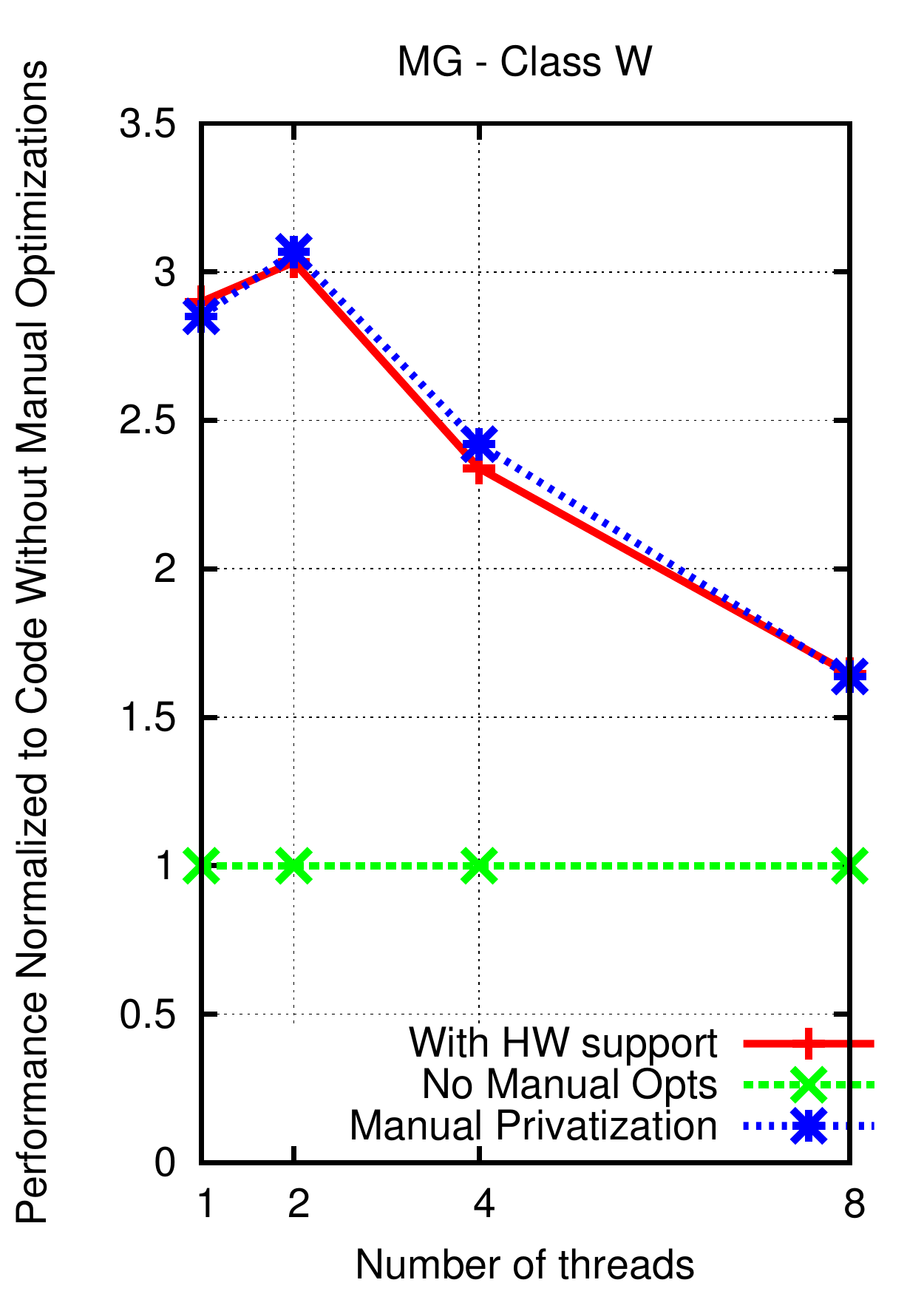}
    }
    \subfloat[Detailed - Execution Time]{
        \includegraphics[width=\halfgraphwidth]{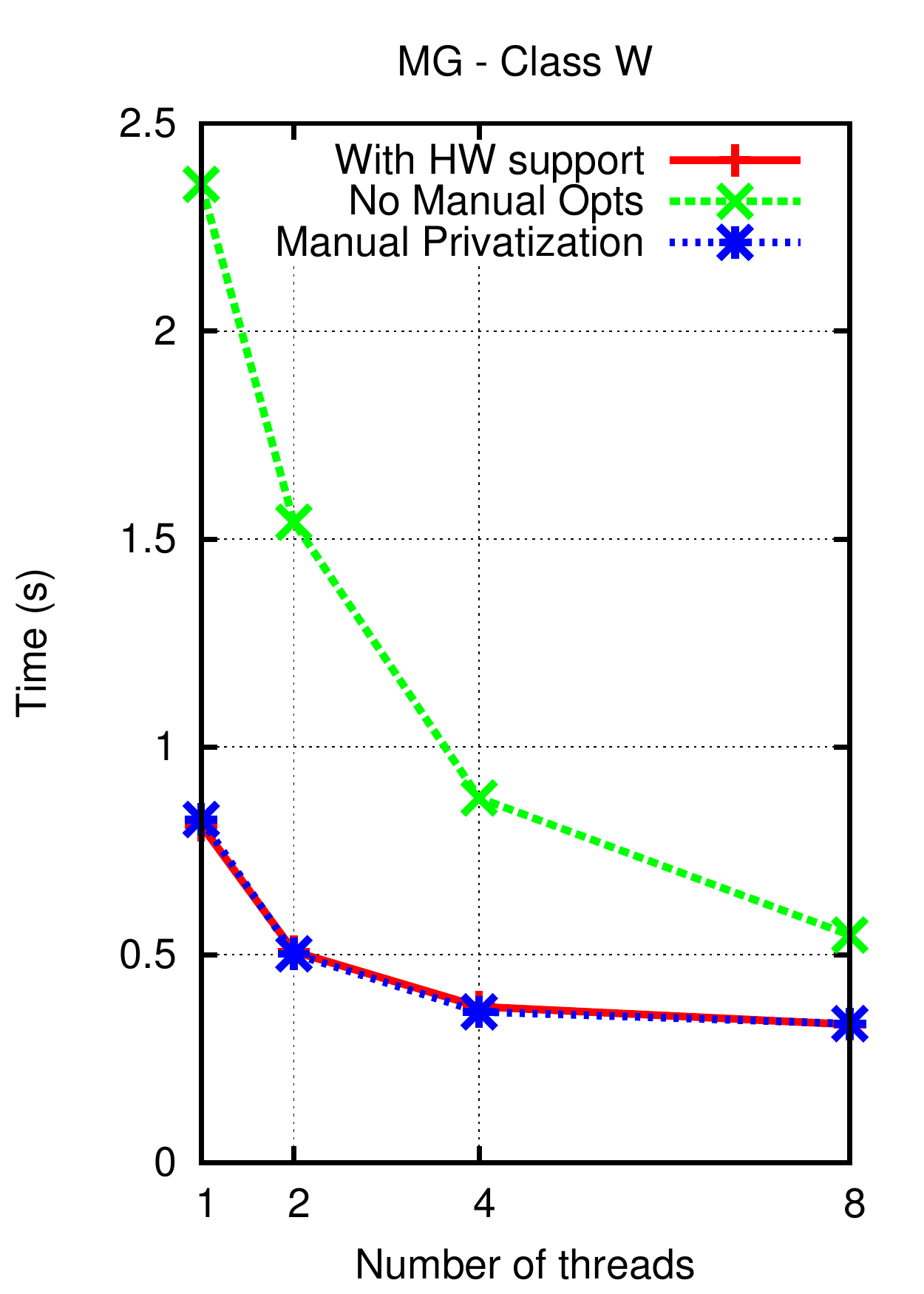}
    }
    \caption{Gem5 : NAS Parallel Benchmark - MG class W}
    \label{fig:results_mg}
\end{figure*}

Gem5 provides different CPU models which covers different architecture
implementations and provides a trade-off between speed and accuracy of the
simulation.
In this work, we used 3 different CPU models: atomic; which is a single
Instruction per Clock (IPC) model, timing; which adds the simulation of the
cache hierarchy and detailed; (also called O3) which simulates a 7-stage,
out-of-order, CPU pipeline.

In order to evaluate the PGAS hardware support, five kernels from the NAS
Parallel Benchmarks \cite{1995_dhb_techrep_npb2}, implemented with UPC
\cite{2002_tel_sc_upcnpb, website_upc_npb}, were used :
\begin{description}
    \item[EP]- Embarrassingly Parallel: generates pairs of Gaussian random
        variates.
    \item[IS]- Integer Sort: Bucket sort of small integers.
    \item[CG]- Conjugate Gradient: approximately computes the small eigenvalues
        of a symmetric positive matrix. Exhibits long-range communication.
    \item[MG]- Multi-Grid: 3D Poisson equation solving using a V-cycle
        multigrid method.
    \item[FT]- Fast Fourier Transform: Solve a partial differential equation
        with a discrete 3D Fast Fourier Transform (FFT).
\end{description}
These benchmarks were implemented with different levels of handmade
optimizations;
we used both the non-optimized version and the manually privatized version of
the benchmarks.
The privatized version was manually optimized in order to replace UPC shared
pointers by private pointers \cite{2001_teg_icpp_upc_benchmarking_issues}.
Due to the very long time needed for such multi-core simulations, only the
relatively small W (workstation) class was used.

Three different results are presented on the graphs: \textit{Without
Manual Optimizations} uses the non hand-optimized NPB kernels with the
unmodified Berkeley compiler with all compiler optimizations enabled,
\textit{Manual Optimization} uses the manually optimized NPB kernels in which
the shared pointers have been replaced by normal C pointers, it is compiled
with the original, unmodified compiler with all the optimization enabled.
Finally, \textit{Without Manual Optimizations, but with HW support} uses the
hardware support with our prototype compiler on the non hand-optimized NPB
kernels.

Figures~\ref{fig:results_ep_atomic}-\ref{fig:results_mg_atomic} present the
runs using the atomic model.
The atomic model of Gem5 being fast, we were able to run the benchmarks with up
to 64 cores (the limit of the BigTsunami architecture).

Figure~\ref{fig:results_ep_atomic} shows that the hardware support does not
provide any performance improvement for the EP kernel; this was expected as EP
does not use any shared pointer in the main loops.

For the CG kernel, not all the shared address incrementation were compiled
using the hardware instructions : the generated code contained 309 shared
address incrementation but 20 of those were using a non-power of 2 element size
(the arrays \texttt{w} and \texttt{w\_tmp} with an element size of 56016);
those incrementations were implemented using software code.
All the other shared pointer manipulation, including 236 loads and stores were
implemented using the hardware instructions.
The CG kernel (Figure~\ref{fig:results_cg_atomic}) runs 2.6 times faster
using the hardware support over the implementation without manual optimizations.
The hardware support is also 17\% faster than the manually optimized version.

For the FT kernel, all the shared pointer manipulations were compiled to
hardware instructions (79 incrementations, and 47 loads and stores).
The FT kernel runs were limited to 16 cores due to the data distribution of the
W class.
The FT kernel performance (Figure~\ref{fig:results_ft_atomic})
is improved 2.3 times without the need of manual optimizations.
The performance also surpasses the manually optimized one by 17\%.
The hardware support can surpass the performance of manually optimized code
because not all the shared pointers were optimized away.
Optimizations often focus on the inner loops and it is not always possible to
remove all shared pointers (due to complex or random access patterns, for
example).

The non-optimized MG kernel code performance is improved by 5.5x (See
Figure~\ref{fig:results_mg_atomic}), but the code is 10\% slower than the
manually optimized version.
Similarly for IS, the base code performance is improved by 3x but with HW
support the code is still 13\% slower than the manually optimized one.
This may be due to some missed optimizations during the C code compilation.
The \texttt{asm} statements for the PGAS stores instructions have been marked
as volatile and changing the memory; this prevents the GCC compiler from moving
the stores around and also forces it to reload data stored in register as
the memory may have changed preventing some optimizations.

Figures~\ref{fig:results_cg}-\ref{fig:results_mg} present the results obtained
using the timing and detailed model. EP (Embarrassingly Parallel) is not
shown as the results are similar to the atomic case since no shared pointers
are used.
The timing model adds caches and memory timing simulation.
The improvements are less substantial, in proportion, as more time is spent
accessing the memory; the single L2 also starts to be a bottleneck with 16
cores.
The detailed memory model introduces an out-of-order processor core.
The number of cores presented for the detailed runs are limited as the
simulator running time becomes very long; multiple days are needed for a
detailed run.

\begin{table*}[!tbp]
    \centering
    \caption{Area cost evaluation for the hardware support}
    \label{tab:areacost}
    \begin{tabular}{|l|r|r|r|r|r|}
        \hline
        \multicolumn{1}{|c|}{\textbf{Configuration}} & \multicolumn{2}{c|}{\textbf{Slice resources}} & \multicolumn{2}{c|}{\textbf{BRAM}} & \textbf{DSP48Es}\\
        \hline
        & \multicolumn{1}{c|}{Registers} & \multicolumn{1}{c|}{LUTs} & \multicolumn{1}{c|}{18kB} & \multicolumn{1}{c|}{36KB} & \\
        \hline
        \hline
        Leon3, 4 cores                         &  $46,718$  &  $59,235$  & 106  &  34  &  16 \\
        Leon3, 4 cores + PGAS hardware support &  $49,325$  &  $62,572$  & 126  &  34  &  24 \\
        \hline
        \hline
        Virtex 6 -  XC6VLX240T                 & $301,440$  & $150,720$  & 832  & 416  & 768 \\
        \hline
        \hline
        Increase                               &   $2,607$  &   $3,337$  & 20          &   0  &  8 \\
        Area increase, \% of base              &  $+5.6\%$  & $+5.6\% $  & $+18.9\%$   &     & $+50.0\%$ \\
        Area \% of Virtex 6          &  $+0.9\%$  & $+2.2\% $  & $+2.4\%$    &     & $+1.0\%$ \\
        \hline
    \end{tabular}
\end{table*}

The detailed model brings more opportunities to reorganize the instructions to
reduce the software overhead to shared address manipulations. However, our
proposed hardware support for PGAS address mapping still provides results that
are comparable or better than the manually optimized codes. In addition, our
proposed hardware mechanism does not require any complex manual tuning of the
code keeping all the productivity advantages of the PGAS model intact.

\subsection{Hardware Implementation Results}

Two micro-benchmarks (vector addition and matrix multiplication) were
implemented to verify the functionality and the performance of the hardware
design.
They were compiled using Berkeley UPC 2.12.1 and GCC 4.4.2 for SPARC with all
the optimizations enabled (-O, -opt for BUPC, -O3 for GCC).
As the support for the extra register file is not implemented in GCC, the
hardware accelerated code was partly written in assembly.

\begin{figure}[!tbp]
    \centering
    \subfloat[Improvement]{
        \includegraphics[width=\halfgraphwidth]{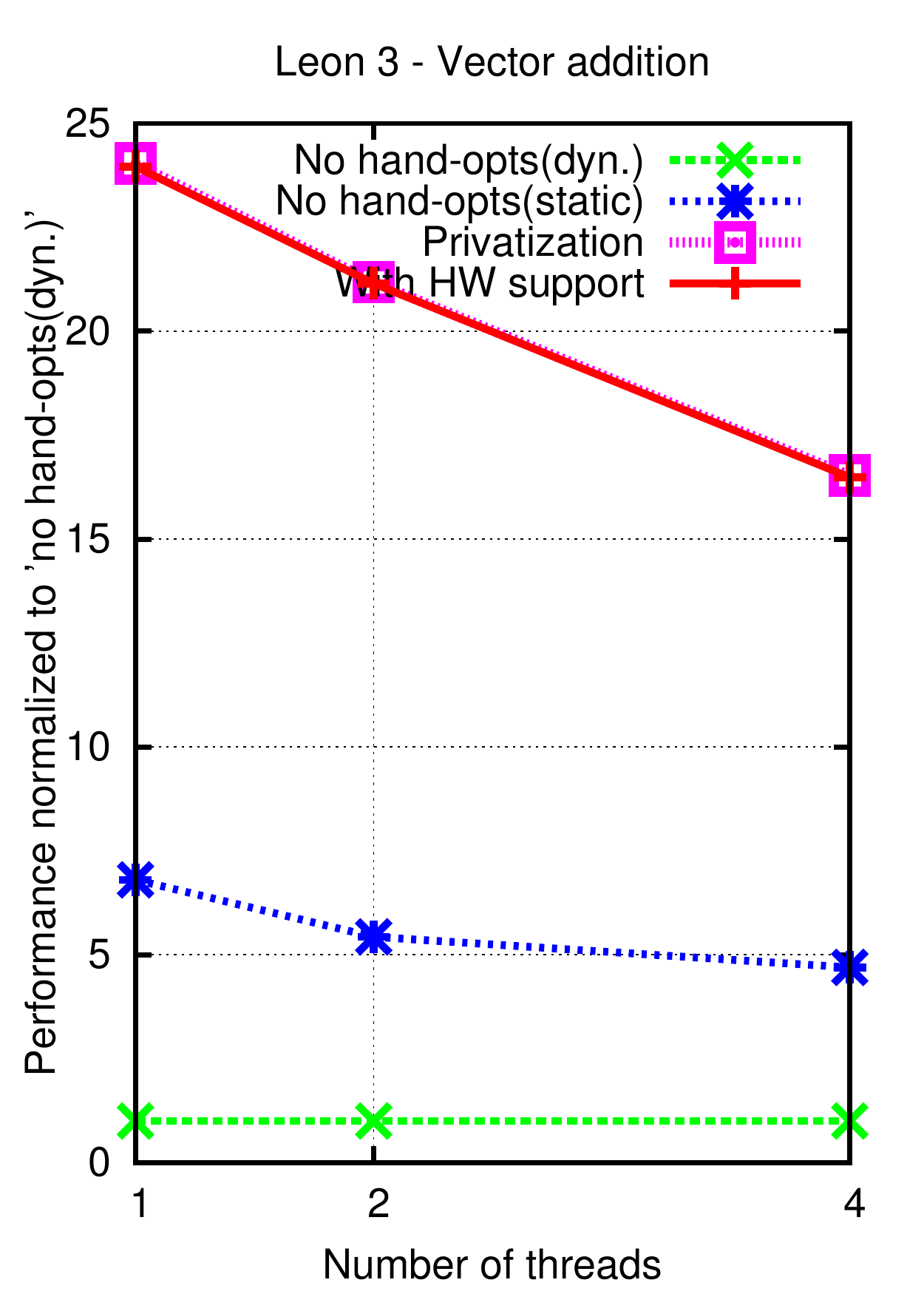}
    }
    \subfloat[Number of cycles]{
        \includegraphics[width=\halfgraphwidth]{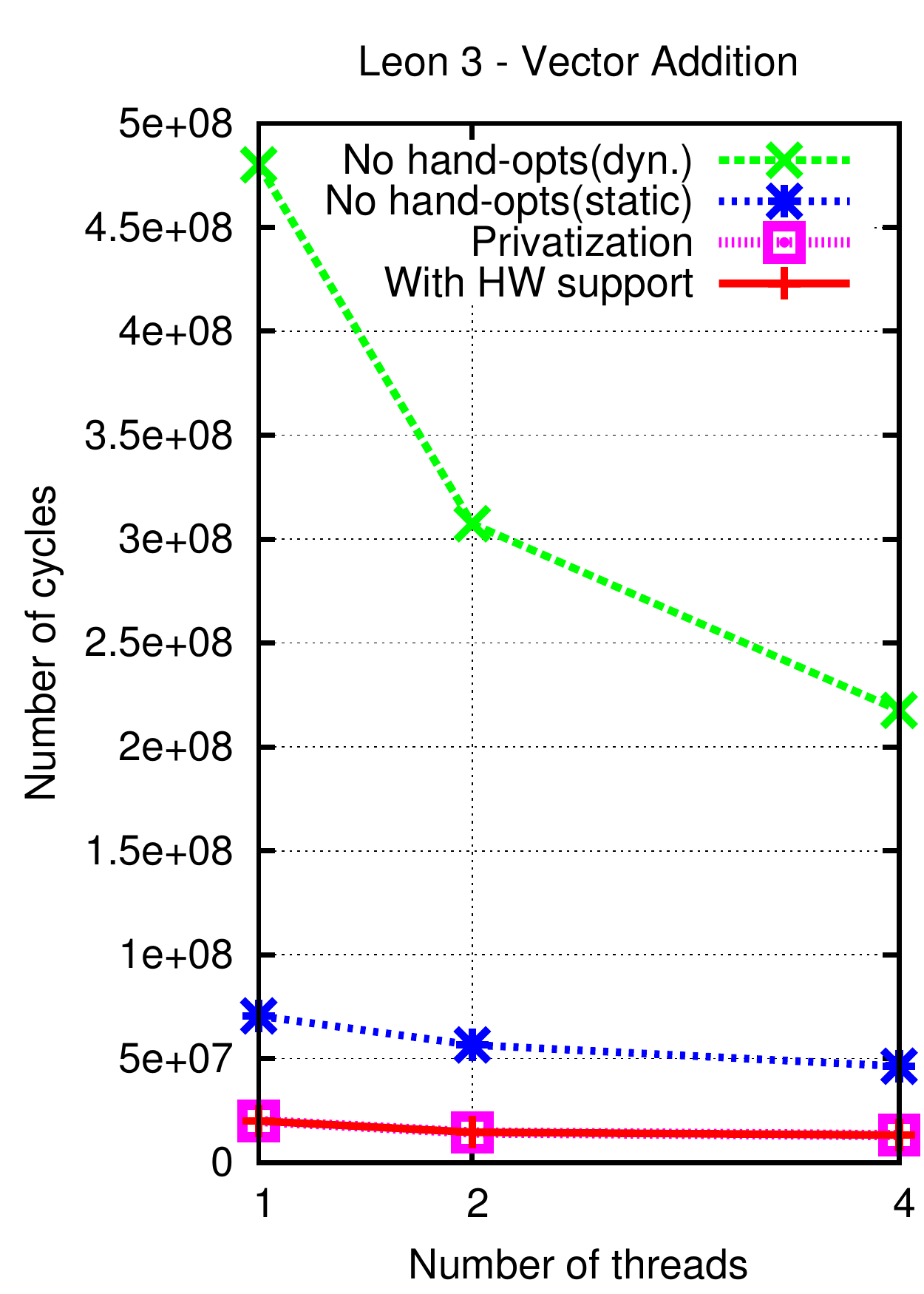}
    }
    \caption{Leon 3 - Vector Addition}
    \label{fig:leon3_vector_addition}
\end{figure}

\begin{figure}[!tbp]
    \centering
    \subfloat[Improvement]{
        \includegraphics[width=\halfgraphwidth]{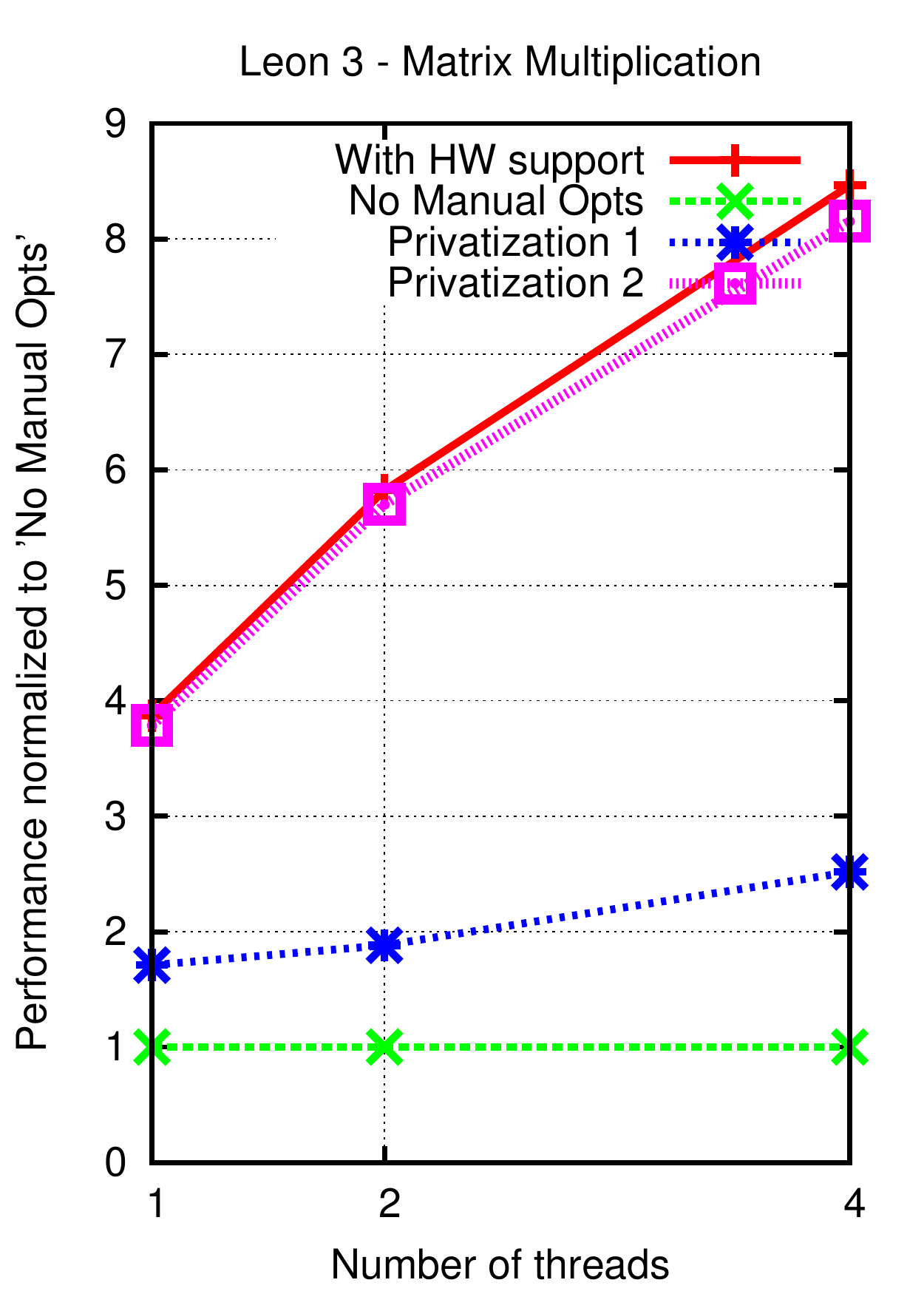}
    }
    \subfloat[Number of cycles]{
        \includegraphics[width=\halfgraphwidth]{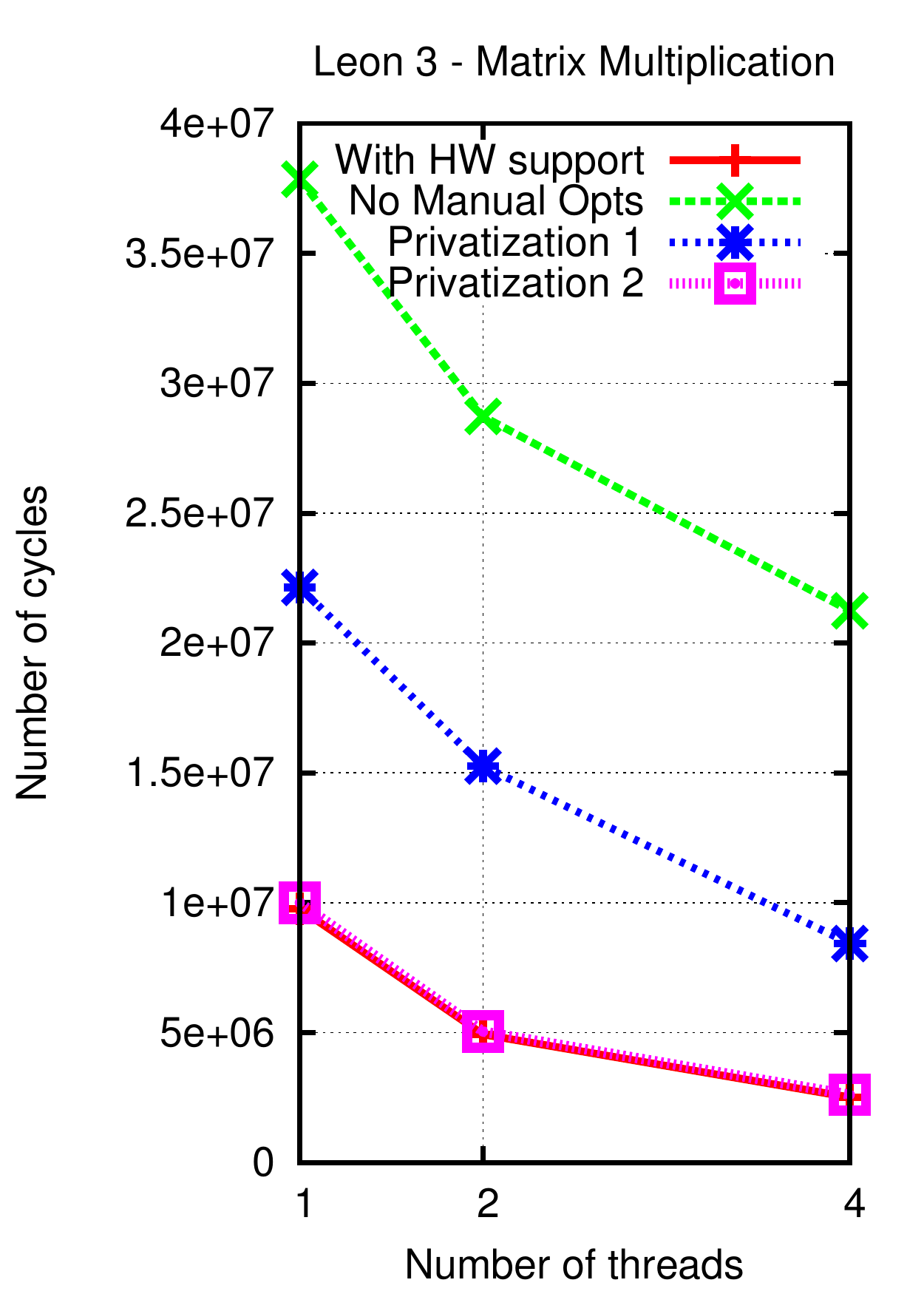}
    }
    \caption{Leon 3 - Matrix Multiplication}
    \label{fig:leon3_matrix_multiplication}
\end{figure}

For the vector addition benchmark, the version of the code
without hand-made optimizations
was compiled both in dynamic mode (without specifying the number of
threads) and in static mode (the number of threads being specified during the
compilation).
In the dynamic mode, the compiler is prevented to optimize the division by
\texttt{threads} during the incrementation of a shared address which reduces
the performance.
Results are shown in Figure~\ref{fig:leon3_vector_addition}.
The code compiled with static runs 5 times faster.
The optimized code to use private pointer or the hardware support both runs 16
times faster (3.5 times faster when compared to the code compiled in the static
mode). The hardware version does not need to be compiled in static mode as
the special register \texttt{threads} can be setup at runtime, allowing the user
to run the same executable with different number of threads.
The performance improvement gets smaller with the number of threads as vector
addition is quickly able to saturate the shared AMBA bus.

The matrix multiplication benchmark was compiled in static mode, i.e. the number
of threads was set at compile time.
Two different levels of manual optimizations were performed: the first one
uses private pointers to access one of the matrix (privatization 1) and the
second one uses a
non-standard UPC extension to be able to access all the matrices with private
pointers.
As seen in Figure~\ref{fig:leon3_matrix_multiplication}, the code with hardware
support matches the performance of the fully optimized version.

More importantly, the FPGA implementation allowed us to evaluate the
chip area needed for the PGAS hardware support.
Table~\ref{tab:areacost} presents the FPGA resources used for a 4 core Leon3
SMP system with and without the hardware support.
Results are both presented in terms of increase compared to the base Leon3
implementation, and in percentage of the FPGA chip used.
The area evaluation is very conservative as we are comparing against a very
simple processor core without floating point support.
Also, adding a new register file (built with BRAM elements) is often not
necessary as the normal register
file can be used to hold shared addresses on 64 bits architectures, as seen in
our Alpha based simulation with Gem5.
The proposed hardware support mechanism for 4 cores utilizes less than 2.4\% of
the overall FPGA chip.

\section{Conclusions}
\label{sec:conclusion}

The PGAS programming model is known for its productivity; however, its
performance can be hindered with the overhead associated with accessing and
traversing its memory model.
Automatic compiler optimizations may help but they are not sufficient for
competitive performance.
Hand tuning of PGAS code can achieve the needed performance levels, but
diminishes the productivity advantage.
In this work, we proposed the addition of a hardware address mapping support
for PGAS.
It was shown through FPGA prototyping that a processor requires only a minimal
increase in the chip area to incorporate this hardware.
In addition, it was shown that this hardware can be availed and used easily by
compilers though simple extensions to the instruction set.
Substantial testing and benchmarking were conducted using a Gem5 full system
simulation as well as FPGA prototyping.
Benchmarking results were based on representative kernels of the well accepted
NAS Parallel Benchmark written with UPC.
Due to the very long time needed for such simulators, only the W class was
used.
In spite of the smaller version of the benchmark, substantial speed up was
achieved.
Larger benchmarks are expected to provide even better scaling results.
The results were consistently comparable to those obtained from hand tuned
code, which demonstrates the power and the productivity of this approach.
The results, using un-optimized code using our proposed hardware support,
achieved up to 5.5 speed up, as compared to the un-optimized code running
without our hardware support.

This work focuses on what we think is presently the biggest impediment of PGAS
languages: the manipulation of shared addresses which create an important
performance penalty even for local accesses.
For future work, we will consider hardware solutions that also allow to further
improve the accesses of remote data across a full system of interconnected node.
This requires extending the PGAS hardware support to the network interface.
We believe that the global solution will be hierarchical to limit the cost of
additional hardware and that the network interface will be able to rely
on shared addresses to quickly locate and communicate with other nodes.

\section*{Acknowledgment}

The authors also wish to acknowledge the Xilinx University Program (XUP) and
the Sun Microsystems OpenSPARC University Program for their hardware and
software donation which has been essential to complete this work.

\bibliographystyle{plain}
\bibliography{refs}

\end{document}